\documentclass{kbsjrnl}
\usepackage{kbsfonts}
\usepackage{multirow}
\newenvironment{descit}[1]{\begin{quote} \textit{#1}}{\end{quote}}

\ifx\undefined\psfig\else \fi

\edef\psfigRestoreAt{\catcode`@=\number\catcode`@\relax}
\catcode`\@=11\relax
\newwrite\@unused
\def\typeout#1{{\let\protect\string\immediate\write\@unused{#1}}}
\def\figurepath{./}

\def\@nnil{\@nil}
\def\@empty{}
\def\@psdonoop#1\@@#2#3{}
\def\@psdo#1:=#2\do#3{\edef\@psdotmp{#2}\ifx\@psdotmp\@empty \else
    \expandafter\@psdoloop#2,\@nil,\@nil\@@#1{#3}\fi}
\def\@psdoloop#1,#2,#3\@@#4#5{\def#4{#1}\ifx #4\@nnil \else
       #5\def#4{#2}\ifx #4\@nnil \else#5\@ipsdoloop #3\@@#4{#5}\fi\fi}
\def\@ipsdoloop#1,#2\@@#3#4{\def#3{#1}\ifx #3\@nnil 
       \let\@nextwhile=\@psdonoop \else
      #4\relax\let\@nextwhile=\@ipsdoloop\fi\@nextwhile#2\@@#3{#4}}
\def\@tpsdo#1:=#2\do#3{\xdef\@psdotmp{#2}\ifx\@psdotmp\@empty \else
    \@tpsdoloop#2\@nil\@nil\@@#1{#3}\fi}
\def\@tpsdoloop#1#2\@@#3#4{\def#3{#1}\ifx #3\@nnil 
       \let\@nextwhile=\@psdonoop \else
      #4\relax\let\@nextwhile=\@tpsdoloop\fi\@nextwhile#2\@@#3{#4}}
\newread\ps@stream
\newif\ifnot@eof       
\newif\if@noisy        
\newif\if@atend        
\newif\if@psfile       
{\catcode`\%=12\global\gdef\epsf@start{
\def\epsf@PS{PS}
\def\epsf@getbb#1{%
\openin\ps@stream=#1
\ifeof\ps@stream\typeout{Error, File #1 not found}\else
   {\not@eoftrue \chardef\other=12
    \def\do##1{\catcode`##1=\other}\dospecials \catcode`\ =10
    \loop
       \if@psfile
	  \read\ps@stream to \epsf@fileline
       \else{
	  \obeyspaces
          \read\ps@stream to \epsf@tmp\global\let\epsf@fileline\epsf@tmp}
       \fi
       \ifeof\ps@stream\not@eoffalse\else
       \if@psfile\else
       \expandafter\epsf@test\epsf@fileline:. \\%
       \fi
          \expandafter\epsf@aux\epsf@fileline:. \\%
       \fi
   \ifnot@eof\repeat
   }\closein\ps@stream\fi}%
\long\def\epsf@test#1#2#3:#4\\{\def\epsf@testit{#1#2}
			\ifx\epsf@testit\epsf@start\else
\typeout{Warning! File does not start with `\epsf@start'.  It may not be a PostScript file.}
			\fi
			\@psfiletrue} 
{\catcode`\%=12\global\let\epsf@percent=
\long\def\epsf@aux#1#2:#3\\{\ifx#1\epsf@percent
   \def\epsf@testit{#2}\ifx\epsf@testit\epsf@bblit
	\@atendfalse
        \epsf@atend #3 . \\%
	\if@atend	
	   \if@verbose{
		\typeout{psfig: found `(atend)'; continuing search}
	   }\fi
        \else
        \epsf@grab #3 . . . \\%
        \not@eoffalse
        \global\no@bbfalse
        \fi
   \fi\fi}%
\def\epsf@grab #1 #2 #3 #4 #5\\{%
   \global\def\epsf@llx{#1}\ifx\epsf@llx\empty
      \epsf@grab #2 #3 #4 #5 .\\\else
   \global\def\epsf@lly{#2}%
   \global\def\epsf@urx{#3}\global\def\epsf@ury{#4}\fi}%
\def\epsf@atendlit{(atend)} 
\def\epsf@atend #1 #2 #3\\{%
   \def\epsf@tmp{#1}\ifx\epsf@tmp\empty
      \epsf@atend #2 #3 .\\\else
   \ifx\epsf@tmp\epsf@atendlit\@atendtrue\fi\fi}

\chardef\letter = 11
\chardef\other = 12

\newif \ifdebug 
\newif\ifc@mpute 
\c@mputetrue 

\let\then = \relax
\def\r@dian{pt }
\let\r@dians = \r@dian
\let\dimensionless@nit = \r@dian
\let\dimensionless@nits = \dimensionless@nit
\def\internal@nit{sp }
\let\internal@nits = \internal@nit
\newif\ifstillc@nverging
\def \Mess@ge #1{\ifdebug \then \message {#1} \fi}

{ 
	\catcode `\@ = \letter
	\gdef \nodimen {\expandafter \n@dimen \the \dimen}
	\gdef \term #1 #2 #3%
	       {\edef \t@ {\the #1}
		\edef \t@@ {\expandafter \n@dimen \the #2\r@dian}%
		\t@rm {\t@} {\t@@} {#3}%
	       }
	\gdef \t@rm #1 #2 #3%
	       {{%
		\count 0 = 0
		\dimen 0 = 1 \dimensionless@nit
		\dimen 2 = #2\relax
		\Mess@ge {Calculating term #1 of \nodimen 2}%
		\loop
		\ifnum	\count 0 < #1
		\then	\advance \count 0 by 1
			\Mess@ge {Iteration \the \count 0 \space}%
			\Multiply \dimen 0 by {\dimen 2}%
			\Mess@ge {After multiplication, term = \nodimen 0}%
			\Divide \dimen 0 by {\count 0}%
			\Mess@ge {After division, term = \nodimen 0}%
		\repeat
		\Mess@ge {Final value for term #1 of 
				\nodimen 2 \space is \nodimen 0}%
		\xdef \Term {#3 = \nodimen 0 \r@dians}%
		\aftergroup \Term
	       }}
	\catcode `\p = \other
	\catcode `\t = \other
	\gdef \n@dimen #1pt{#1} 
}

\def \Divide #1by #2{\divide #1 by #2} 

\def \Multiply #1by #2
       {{
	\count 0 = #1\relax
	\count 2 = #2\relax
	\count 4 = 65536
	\Mess@ge {Before scaling, count 0 = \the \count 0 \space and
			count 2 = \the \count 2}%
	\ifnum	\count 0 > 32767 
	\then	\divide \count 0 by 4
		\divide \count 4 by 4
	\else	\ifnum	\count 0 < -32767
		\then	\divide \count 0 by 4
			\divide \count 4 by 4
		\else
		\fi
	\fi
	\ifnum	\count 2 > 32767 
	\then	\divide \count 2 by 4
		\divide \count 4 by 4
	\else	\ifnum	\count 2 < -32767
		\then	\divide \count 2 by 4
			\divide \count 4 by 4
		\else
		\fi
	\fi
	\multiply \count 0 by \count 2
	\divide \count 0 by \count 4
	\xdef \product {#1 = \the \count 0 \internal@nits}%
	\aftergroup \product
       }}

\def\r@duce{\ifdim\dimen0 > 90\r@dian \then   
		\multiply\dimen0 by -1
		\advance\dimen0 by 180\r@dian
		\r@duce
	    \else \ifdim\dimen0 < -90\r@dian \then  
		\advance\dimen0 by 360\r@dian
		\r@duce
		\fi
	    \fi}

\def\Sine#1%
       {{%
	\dimen 0 = #1 \r@dian
	\r@duce
	\ifdim\dimen0 = -90\r@dian \then
	   \dimen4 = -1\r@dian
	   \c@mputefalse
	\fi
	\ifdim\dimen0 = 90\r@dian \then
	   \dimen4 = 1\r@dian
	   \c@mputefalse
	\fi
	\ifdim\dimen0 = 0\r@dian \then
	   \dimen4 = 0\r@dian
	   \c@mputefalse
	\fi
	\ifc@mpute \then
		\divide\dimen0 by 180
		\dimen0=3.141592654\dimen0
		\dimen 2 = 3.1415926535897963\r@dian 
		\divide\dimen 2 by 2 
		\Mess@ge {Sin: calculating Sin of \nodimen 0}%
		\count 0 = 1 
		\dimen 2 = 1 \r@dian 
		\dimen 4 = 0 \r@dian 
		\loop
			\ifnum	\dimen 2 = 0 
			\then	\stillc@nvergingfalse 
			\else	\stillc@nvergingtrue
			\fi
			\ifstillc@nverging 
			\then	\term {\count 0} {\dimen 0} {\dimen 2}%
				\advance \count 0 by 2
				\count 2 = \count 0
				\divide \count 2 by 2
				\ifodd	\count 2 
				\then	\advance \dimen 4 by \dimen 2
				\else	\advance \dimen 4 by -\dimen 2
				\fi
		\repeat
	\fi		
			\xdef \sine {\nodimen 4}%
       }}

\def\Cosine#1{\ifx\sine\UnDefined\edef\Savesine{\relax}\else
		             \edef\Savesine{\sine}\fi
	{\dimen0=#1\r@dian\multiply\dimen0 by -1
	 \advance\dimen0 by 90\r@dian
	 \Sine{\nodimen 0}
	 \xdef\cosine{\sine}
	 \xdef\sine{\Savesine}}}	      

\def\psdraft{
	\def\@psdraft{0}
}
\def\psfull{
	\def\@psdraft{100}
}

\psfull

\newif\if@draftbox
\def\psnodraftbox{
	\@draftboxfalse
}
\@draftboxtrue

\newif\if@prologfile
\newif\if@postlogfile
\def\pssilent{
	\@noisyfalse
}
\def\psnoisy{
	\@noisytrue
}
\psnoisy
\newif\if@bbllx
\newif\if@bblly
\newif\if@bburx
\newif\if@bbury
\newif\if@height
\newif\if@width
\newif\if@rheight
\newif\if@rwidth
\newif\if@angle
\newif\if@clip
\newif\if@verbose
\newif\if@scale
\def\@p@@sclip#1{\@cliptrue}

\def\@p@@sfile#1{\def\@p@sfile{null}%
	        \openin1=#1
		\ifeof1\closein1%
		       \openin1=\figurepath#1
			\ifeof1\typeout{Error, File #1 not found}
			   \if@bbllx\if@bblly\if@bburx\if@bbury
			      \def\@p@sfile{#1}%
			   \fi\fi\fi\fi
			\else\closein1
			    \edef\@p@sfile{\figurepath#1}%
                        \fi%
		 \else\closein1%
		       \def\@p@sfile{#1}%
		 \fi}
\def\@p@@sfigure#1{\def\@p@sfile{null}%
	        \openin1=#1
		\ifeof1\closein1%
		       \openin1=\figurepath#1
			\ifeof1\typeout{Error, File #1 not found}
			   \if@bbllx\if@bblly\if@bburx\if@bbury
			      \def\@p@sfile{#1}%
			   \fi\fi\fi\fi
			\else\closein1
			    \def\@p@sfile{\figurepath#1}%
                        \fi%
		 \else\closein1%
		       \def\@p@sfile{#1}%
		 \fi}

\def\@p@@sbbllx#1{
		\@bbllxtrue
		\dimen100=#1
		\edef\@p@sbbllx{\number\dimen100}
}
\def\@p@@sbblly#1{
		\@bbllytrue
		\dimen100=#1
		\edef\@p@sbblly{\number\dimen100}
}
\def\@p@@sbburx#1{
		\@bburxtrue
		\dimen100=#1
		\edef\@p@sbburx{\number\dimen100}
}
\def\@p@@sbbury#1{
		\@bburytrue
		\dimen100=#1
		\edef\@p@sbbury{\number\dimen100}
}
\def\@p@@sheight#1{
		\@heighttrue
		\dimen100=#1
   		\edef\@p@sheight{\number\dimen100}
}
\def\@p@@swidth#1{
		\@widthtrue
		\dimen100=#1
		\edef\@p@swidth{\number\dimen100}
}
\def\@p@@srheight#1{
		\@rheighttrue
		\dimen100=#1
		\edef\@p@srheight{\number\dimen100}
}
\def\@p@@srwidth#1{
		\@rwidthtrue
		\dimen100=#1
		\edef\@p@srwidth{\number\dimen100}
}
\def\@p@@sangle#1{
		\@angletrue
		\edef\@p@sangle{#1} 
}
\def\@p@@ssilent#1{ 
		\@verbosefalse
}
\def\@p@@sscale#1{
		\def\@p@scale{#1}
		\@scaletrue
}
\def\@p@@sprolog#1{\@prologfiletrue\def\@prologfileval{#1}}
\def\@p@@spostlog#1{\@postlogfiletrue\def\@postlogfileval{#1}}
\def\@cs@name#1{\csname #1\endcsname}
\def\@setparms#1=#2,{\@cs@name{@p@@s#1}{#2}}
\def\ps@init@parms{
		\@bbllxfalse \@bbllyfalse
		\@bburxfalse \@bburyfalse
		\@heightfalse \@widthfalse
		\@rheightfalse \@rwidthfalse
		\@scalefalse
		\def\@p@sbbllx{}\def\@p@sbblly{}
		\def\@p@sbburx{}\def\@p@sbbury{}
		\def\@p@sheight{}\def\@p@swidth{}
		\def\@p@srheight{}\def\@p@srwidth{}
		\def\@p@sangle{0}
		\def\@p@sfile{}
		\def\@p@scost{10}
		\def\@sc{}
		\@prologfilefalse
		\@postlogfilefalse
		\@clipfalse
		\if@noisy
			\@verbosetrue
		\else
			\@verbosefalse
		\fi
}
\def\parse@ps@parms#1{
	 	\@psdo\@psfiga:=#1\do
		   {\expandafter\@setparms\@psfiga,}}
\newif\ifno@bb
\def\bb@missing{
	\if@verbose{
		\typeout{psfig: searching \@p@sfile \space  for bounding box}
	}\fi
	\no@bbtrue
	\epsf@getbb{\@p@sfile}
        \ifno@bb \else \bb@cull\epsf@llx\epsf@lly\epsf@urx\epsf@ury\fi
}	
\def\bb@cull#1#2#3#4{
	\dimen100=#1 bp\edef\@p@sbbllx{\number\dimen100}
	\dimen100=#2 bp\edef\@p@sbblly{\number\dimen100}
	\dimen100=#3 bp\edef\@p@sbburx{\number\dimen100}
	\dimen100=#4 bp\edef\@p@sbbury{\number\dimen100}
	\no@bbfalse
}

\newdimen\p@intvaluex
\newdimen\p@intvaluey
\newdimen\@ffsetvalue
\newdimen\x@ffsetvalue
\newdimen\y@ffsetvalue

\def\compute@offset#1#2{{\dimen0=#1 sp\dimen1=#2 sp
			\advance\dimen1 by -\dimen0
			\dimen1=\sine\dimen1
			\dimen0=\cosine\dimen1
			\ifdim\dimen0<0sp \dimen1=0sp \fi
			\global\@ffsetvalue=\dimen1}}

\def\rotate@#1#2{{\dimen0=#1 sp\dimen1=#2 sp
		  \global\p@intvaluex=\cosine\dimen0
		  \dimen3=\sine\dimen1
		  \global\advance\p@intvaluex by -\dimen3
		  \global\p@intvaluey=\sine\dimen0
		  \dimen3=\cosine\dimen1
		  \global\advance\p@intvaluey by \dimen3
		  }}
\def\compute@bb{
		\no@bbfalse
		\if@bbllx \else \no@bbtrue \fi
		\if@bblly \else \no@bbtrue \fi
		\if@bburx \else \no@bbtrue \fi
		\if@bbury \else \no@bbtrue \fi
		\ifno@bb \bb@missing \fi
		\ifno@bb \typeout{FATAL ERROR: no bb supplied or found}
			\no-bb-error
		\fi
		\if@angle 
			\Sine{\@p@sangle}\Cosine{\@p@sangle}
			\compute@offset{\@p@sbblly}{\@p@sbbury}
			\x@ffsetvalue=\@ffsetvalue
			\compute@offset{\@p@sbburx}{\@p@sbbllx}
			\y@ffsetvalue=\@ffsetvalue

			\rotate@{\@p@sbbllx}{\@p@sbblly}
			\advance\p@intvaluex by -\x@ffsetvalue
			\advance\p@intvaluey by -\y@ffsetvalue
			\edef\@p@sbbllx{\number\p@intvaluex}
			\edef\@p@sbblly{\number\p@intvaluey}

			\rotate@{\@p@sbburx}{\@p@sbbury}
			\advance\p@intvaluex by \x@ffsetvalue
			\advance\p@intvaluey by \y@ffsetvalue
			\edef\@p@sbburx{\number\p@intvaluex}
			\edef\@p@sbbury{\number\p@intvaluey}
			{
			 \count0=\@p@sbbllx \count1=\@p@sbblly
		 	 \count2=\@p@sbburx \count3=\@p@sbbury
			 \dimen0=\@p@sbbllx sp\dimen1=\@p@sbblly sp
		 	 \dimen2=\@p@sbburx sp\dimen3=\@p@sbbury sp
			 \dimen203=\dimen2 \advance\dimen203 by -\dimen0
			 \dimen204=\dimen3 \advance\dimen204 by -\dimen1
			 \ifdim\dimen203<0sp 
			      \count203=\count2 \count2=\count0 
			      \count0=\count203 
			      \global\edef\@p@sbbllx{\number\count0}
			      \global\edef\@p@sbburx{\number\count2}
			 \fi
			 \ifdim\dimen204<0sp 
			       \count204=\count3
			       \count3=\count1
			       \count1=\count204
			       \global\edef\@p@sbblly{\number\count1}
			       \global\edef\@p@sbbury{\number\count3}
			 \fi
			}
		\fi
		\count203=\@p@sbburx
		\count204=\@p@sbbury
		\advance\count203 by -\@p@sbbllx
		\advance\count204 by -\@p@sbblly
		\edef\@bbw{\number\count203}
		\edef\@bbh{\number\count204}
}
\def\in@hundreds#1#2#3{\count240=#2 \count241=#3
		     \count100=\count240	
		     \divide\count100 by \count241
		     \count101=\count100
		     \multiply\count101 by \count241
		     \advance\count240 by -\count101
		     \multiply\count240 by 10
		     \count101=\count240	
		     \divide\count101 by \count241
		     \count102=\count101
		     \multiply\count102 by \count241
		     \advance\count240 by -\count102
		     \multiply\count240 by 10
		     \count102=\count240	
		     \divide\count102 by \count241
		     \count200=#1\count205=0
		     \count201=\count200
			\multiply\count201 by \count100
		 	\advance\count205 by \count201
		     \count201=\count200
			\divide\count201 by 10
			\multiply\count201 by \count101
			\advance\count205 by \count201
		     \count201=\count200
			\divide\count201 by 100
			\multiply\count201 by \count102
			\advance\count205 by \count201
		     \edef\@result{\number\count205}
}
\def\@ScaleInHundreds#1{
		\in@hundreds{#1}{\@p@scale}{100}
		\edef#1{\@result}
}
\def\compute@wfromh{
		\in@hundreds{\@p@sheight}{\@bbw}{\@bbh}
		\edef\@p@swidth{\@result}
}
\def\compute@hfromw{
		\in@hundreds{\@p@swidth}{\@bbh}{\@bbw}
		\edef\@p@sheight{\@result}
}
\def\compute@handw{
		\if@height 
			\if@width
			\else
				\compute@wfromh
			\fi
		\else 
			\if@width
				\compute@hfromw
			\else
				\edef\@p@sheight{\@bbh}
				\edef\@p@swidth{\@bbw}
			\fi
		\fi
}
\def\compute@resv{
		\if@rheight \else \edef\@p@srheight{\@p@sheight} \fi
		\if@rwidth \else \edef\@p@srwidth{\@p@swidth} \fi
}
\def\compute@sizes{
	\compute@bb
	\compute@handw
	\compute@resv
}
\def\psfig#1{\vbox {
	\ps@init@parms
	\parse@ps@parms{#1}
	\compute@sizes
	\if@scale
                \if@verbose
                        \typeout{psfig: scaling by \@p@scale}
                \fi
                \@ScaleInHundreds{\@p@swidth}
                \@ScaleInHundreds{\@p@sheight}
                \@ScaleInHundreds{\@p@srwidth}
                \@ScaleInHundreds{\@p@srheight}
        \fi
	\ifnum\@p@scost<\@psdraft{
		\if@verbose{
			\typeout{psfig: including \@p@sfile \space }
		}\fi
		\special{ps::[begin] 	\@p@swidth \space \@p@sheight \space
				\@p@sbbllx \space \@p@sbblly \space
				\@p@sbburx \space \@p@sbbury \space
				startTexFig \space }
		\if@angle
			\special {ps:: \@p@sangle \space rotate \space} 
		\fi
		\if@clip{
			\if@verbose{
				\typeout{(clip)}
			}\fi
			\special{ps:: doclip \space }
		}\fi
		\if@prologfile
		    \special{ps: plotfile \@prologfileval \space } \fi
		\special{ps: plotfile \@p@sfile \space }
		\if@postlogfile
		    \special{ps: plotfile \@postlogfileval \space } \fi
		\special{ps::[end] endTexFig \space }
		\vbox to \@p@srheight true sp{
			\hbox to \@p@srwidth true sp{
				\hss
			}
		\vss
		}
	}\else{
		\if@draftbox{		
			\hbox{\fbox{\vbox to \@p@srheight true sp{
			\vss
			\hbox to \@p@srwidth true sp{ \hss \@p@sfile \hss }
			\vss
			}}}
		}\else{
			\vbox to \@p@srheight true sp{
			\vss
			\hbox to \@p@srwidth true sp{\hss}
			\vss
			}
		}\fi

	}\fi
}}
\def\psglobal{\typeout{psfig: PSGLOBAL is OBSOLETE; use psprint -m instead}}
\psfigRestoreAt

\newif\ifpdf
\ifx\pdfoutput\undefined
  \pdffalse
\else
  \pdfoutput=1
  \pdftrue
\fi

\ifpdf
  \usepackage[pdftex]{graphicx}
  \usepackage[pdftex]{color}
  \DeclareGraphicsExtensions{.pdf,.png,.jpg}
\else
  \usepackage[dvips]{graphicx}
  \usepackage[dvips]{color}
  \DeclareGraphicsExtensions{.eps,.epsi,.ps}
\fi

\begin{document}

\authorrunninghead{Perugini, Gon{\c{c}}alves and Fox}
\titlerunninghead{A Connection-Centric Survey of Recommender Systems Research}

\setcounter{page}{1}

\begin{article}

\title{A Connection-Centric Survey of\\Recommender Systems Research}

\authors{Saverio Perugini}
\email{sperugin@cs.vt.edu}
\affil{Department of Computer Science, Virginia Tech, Blacksburg, VA 24061}

\authors{Marcos Andr{\'{e}} Gon{\c{c}}alves}
\email{mgoncalv@cs.vt.edu}
\affil{Department of Computer Science, Virginia Tech, Blacksburg, VA 24061}

\authors{Edward A. Fox}
\email{fox@cs.vt.edu}
\affil{Department of Computer Science, Virginia Tech, Blacksburg, VA 24061}

\abstract{Recommender systems attempt to reduce information overload
and retain customers by selecting a subset of items from a
universal set based on user preferences.
While research in recommender systems
grew out of information retrieval and filtering, the topic has
steadily advanced into a legitimate and challenging research area of its own.
Recommender systems have traditionally been studied from a content-based
filtering vs. collaborative design perspective.
Recommendations, however, are
not delivered within a vacuum, but rather cast within an informal
community of users
and social context. Therefore, ultimately all recommender systems
make {\em connections} among people
and thus should be surveyed from such a perspective.
This viewpoint is under-emphasized in the recommender systems
literature.  We therefore take
a {\em connection-oriented} viewpoint toward recommender systems research.
We posit that recommendation has an inherently social element and
is ultimately intended to connect people
either directly as a result of explicit user modeling or
indirectly through the discovery of relationships implicit in extant data.
Thus, recommender systems
are characterized by how they model users to bring people together:
explicitly or implicitly.
Finally, user modeling and the connection-centric viewpoint 
raise broadening and social issues---such as evaluation, targeting, and
privacy and trust---which we also briefly address.}


\begin{descit}{}
\noindent
``What information consumes is rather obvious: it consumes the attention of
its recipients. Hence a wealth of information creates a poverty of attention,
and a need to allocate that attention efficiently among the overabundance of
information sources that might consume it.''

\flushright{Herbert~A.~Simon}
\end{descit}

\section{Introduction}

The advent of the WWW and concomitant increase in information
available online has caused information overload and
ignited research in recommender systems.
By selecting a subset of items from a universal set based on user preferences,
recommender systems attempt to reduce information overload and retain customers.
Examples of systems include top-$N$ lists, book~\cite{bookContent}
and movie~\cite{movie-content} recommenders, advanced
search engines~\cite{miningLinkStr}, and intelligent
avatars~\cite{webCompanions}.
The benefits of recommendation are most salient in
voluminous and ephemeral domains~(e.g.,~news)
and include `predictive utility'~\cite{GroupLens}, the
value of a recommendation as advice given
prior to investing time, energy, and in most
cases, money in consuming a product.
Recommender systems harness techniques which develop a model of
user preferences to predict future ratings of artifacts.
The underlying algorithms to realize recommendation
range from keyword matching~\cite{keyword} to sophisticated data mining
of customer profiles~\cite{miningPers}.
Recommender systems are now widely believed to be critical to
sustaining the Internet economy~\cite{InformationRules}.

Four main dimensions have been identified to help in the
study recommender systems: how the system
is~(i) modeled and designed~(i.e.,~are recommendations content-based
or collaborative?),~(ii) targeted~(to an individual, group, or topic),~(iii)
built, and~(iv) maintained~(online vs. offline)~\cite{BatulThesis}.
Recommender systems are typically studied based on the approach to
modeling, of
which the most popular~(and over-emphasized) is
{\em content-based filtering}~(i.e.,~recommend items similar to those
I have liked in
the past; e.g., `Since you liked {\em The Little Lisper}, you also might be
interested in {\em The Little Schemer}.')~\cite{bookContent}.
An alternate approach to modeling is
{\em collaborative filtering}~(i.e.,~recommend items that users,
whose tastes are similar to mine, have liked; e.g., `Linus and Lucy like
{\em Sleepless in Seattle}.  Linus likes {\em You've Got Mail}.
Lucy also might like {\em You've Got Mail}.')~\cite{cf}.
For example,
Terveen and Hill survey content-based and collaborative filtering systems
in a human-computer interaction~(HCI) context~\cite{HCIbookRS}.
Others classify recommender
systems from a business-oriented perspective~\cite{recEComm}, often based
on how they are {\it built}.
For instance, Schafer, Konstan, and Riedl survey 
recommender systems in e-commerce
based on interface, technology, and recommendation discovery~\cite{recEComm}.
These researchers also cast these aspects of
recommenders in a two-dimensional space of
recommendation lifetime~(ephemeral
vs. persistent) and level of automation~(manual vs. automatic) which
connects with how they are {\it maintained}.
Recommender systems, however, have an inherently social element and
ultimately bring
people together---a viewpoint under-emphasized in the literature---and
therefore should be surveyed from this perspective.

Consider that the process of recommendation in a `brick and mortar' setting
is inherently dependent on knowledge of personal taste.
For example, in a restaurant with a friend, the following dialog might
arise:
`The menu looks enticing.  Since you are a returning patron,
what do you recommend?'  `Well, since you like spicy dishes,
you may enjoy the chili chicken curry.'  A mutually reinforcing
dynamic ensues.  The recommender's personal knowledge of
her friend's interests are incorporated into the recommendation
process.  Conversely, after a recommendation is made, the
recipient's personal knowledge of the recommender's reputation helps
him evaluate the recommendation.
Recommender systems attempt to emulate and
automate this naturally social process.
This seemingly simple example speaks volumes about the process of
making recommendations.  Not only does a recommender system have an
underlying social
element, but its effectiveness
is predicated upon its representation of the recipient.
Therefore, recommender systems involve user modeling, which
includes developing a representation of user preferences and interests.
User models can be constructed by explicitly soliciting feedback
(e.g., asking the user to rate products or services)~\cite{GroupLens}
or gleaning implicit declarations of interest~(e.g.,~through monitoring
usage)~\cite{PHOAKS}.

\begin{figure}[t]
\centering
\begin{tabular}{c}
\includegraphics[width=11.0cm,height=8.25cm]{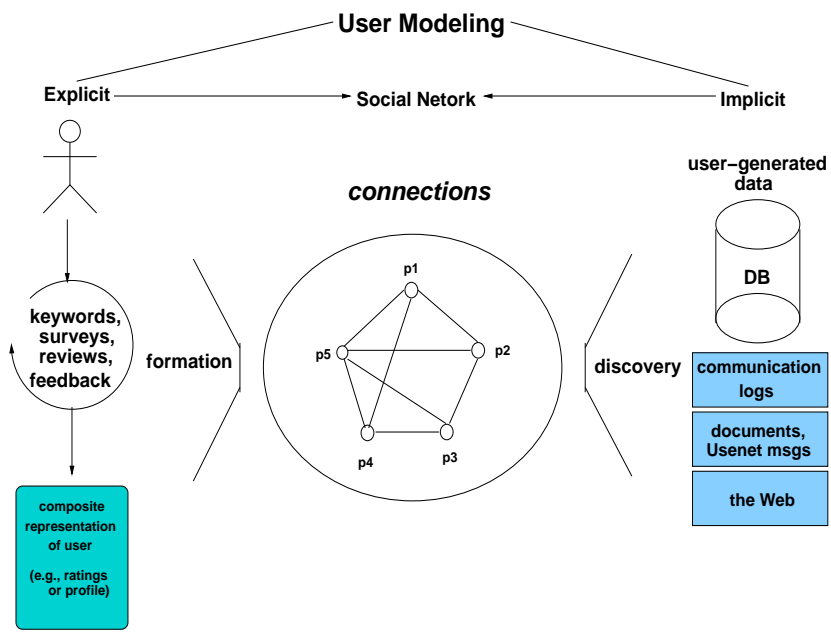}
\end{tabular}
\caption{A connection-centric view of recommendation as
bringing people together into a social network~(center).
(left) {\em Formation} of a social
network by {\em explicitly} collecting ratings or
profiles.  (right) Identification and {\em discovery} of
a network by exposing
self-organizing communities {\em implicit} in user-generated data such as
communication or web logs.
Although not illustrated explicitly, these two approaches may be combined.}
\label{overview}
\end{figure}

User modeling is directed toward developing a basis to compute overlap,
and ultimately is conducted to make {\it connections}
among people to drive recommendation.  Thus, once enough users are engaged
and modeled 
to sufficiently sustain a system,
connections~(recommendations) can be made.  Recommendations, thus, are
not delivered within a vacuum, but rather cast within an
`informal [community] of collaborators, colleagues,
or friends'~\cite{ReferralWeb}, known as a
social network~\cite{AdvancesSNA}.
Explicit user modeling~(and correlating the resulting ratings) then
can be seen as directed toward forming such connected~(community) graph
components.  Collecting
implicit declarations of preference also can be viewed as
directed toward inducing social networks.  This is analogous to
techniques to discover existing social networks
from patterns embedded in interaction~(transaction) data.
Therefore an extension to implicit user modeling and an
approach toward a basis to compute recommendations
entails directly exposing these self-organizing and self-maintaining
social structures.  Since social networks model social processes, these
informal communities with shared interests are implicit
in data generated automatically by electronic communications.
This extension is corroborated by a recent
trend toward exploring and exploiting connections of social processes
in graph representations of self-organizing structures, such as the web,
as a viable and increasingly popular way to satisfy information-seeking and
recommendation-oriented
goals~\cite{SIGIR2003keynote,WebGraph,WebStrucScience}.
This less invasive approach not only
supersedes the need to explicitly model users individually, but also
results in more natural, reflective,
and fertile organizations for recommendation.
Exploration of identified
existing social networks fosters the discovery of serendipitous
connections~\cite{socialNet}, social referrals~\cite{ReferralWeb}, and
\mbox{cyber-communities}~\cite{trawling}---and hence
offers many opportunities for recommendation.
The use of social networks has expanded to many diverse application
domains such as movie recommendation~\cite{jumpingConnections},
digital libraries~\cite{bioDL}, and community-based service
location~\cite{commBasedServiceLoc}.

This connection-oriented viewpoint
and these two ways of realizing it provide the basis for this survey.  
We posit that recommendation has an inherently social element and
is ultimately concerned with connecting people
either directly as a result of explicit user modeling or indirectly through the
discovery of relationships implicit in existing data
(see Fig.~\ref{overview}).  We make connection-based distinctions.
Systems are characterized by how they model users to bring people together:
explicitly or implicitly.  The goal then
of a recommender system is to bring as many people together as possible,
which also suggests a novel evaluation criterion~(e.g.,~algorithm
$A$ connects $x$ individuals while algorithm
$B$ connects $y$)~\cite{jumpingConnections}.
Thus, while
Amazon may make better book recommendations than Barnes and Noble, if they
arrive at connected user components in the same manner, then
in this survey they would be considered equivalent.

\subsubsection*{Reader's Guide}

The balance of this survey is organized as follows.
Section~\ref{history} presents an historical perspective of recommender systems
and outlines their evolution from IR.
Section~\ref{user-modeling} showcases techniques to
explicitly model users for social network formation while
Section~\ref{implicit} describes approaches to identifying
existing communities to explore for recommendation.
The relative lengths of these two sections
reflect the emphasis each places on connections.  Implicit modeling
approaches and resulting systems make social networks 
salient and thus are treated in greater detail.
User modeling and the connection-centric viewpoint
raise broadening and social issues, such as
evaluation, targeting,
and privacy and trust, which we cursorily address in Section~\ref{broadening}.
We identify various opportunities for future research in Section~\ref{future}.

\section{A Chequered History}
\label{history}

While Amazon.com~\cite{Amazon}, a pioneer in the \mbox{e-commerce}
revolution, \mbox{spear-headed} a movement
toward recommenders and was instrumental
in bringing such systems to critical mass,
recommender systems research is a result of 
a series of shifts in information systems (IS) research.
In the 1970s a great deal of IS
research was focused on IR.  In this era \mbox{Salton} and his
students developed
the \mbox{vector-space} model~\cite{vector-space} and
the SMART system~\cite{relevanceFeedback}.
Researchers modeled IR systems with large sparse (and anti-symmetric)
term-document matrices which
permitted document similarity to be measured
by the cosine of the angle between vectors in a multi-dimensional
space.  Precision and recall became the two quintessential IR
metrics~\cite{SaltonIRBook}.
The emphasis of such research and systems was on satisfying
short-term information-seeking goals by retrieving
information deemed relevant to queries.  IR research flourished
in this period and many
supportive techniques such as relevance feedback~\cite{relevanceFeedback}
were developed, demonstrating qualified success.

As the end of the 1970s drew near, electronic information become
more abundant.  The 1980s brought a rapid proliferation of
information due to desktop computers and
applications such as
word processors and spreadsheets.  In addition, the
introduction of e-mail into the mainstream further exasperated the
copious amounts of text residing in computers~(termed `electronic junk'
by \mbox{Denning}~\cite{electronicJunk}).
The new found ease of information generation ignited a shift in 
IS research initiatives.  Researchers began to focus on 
removing irrelevant information rather than retrieving relevant
information.  Information categorization, routing, and filtering became
of immediate importance.  This first shift spawned an {\em information
filtering} thread.

In 1991 Bellcore hosted a workshop on information filtering (IF) which
lead to the December 1992 {\it Communications of the ACM}
special issue on the topic~\cite{if}.
In this issue \mbox{Belkin} and \mbox{Croft}
compared and contrasted IF and IR~\cite{irif}.
While IR entails returning relevant information in response to
short-term information-seeking goals via
requests such as queries,
information filtering involves removing persistent and 
irrelevant information over a long period of time.
Information filtering systems model document features in user
profiles~\cite{bookContent}, which
replaced terms in a modeling matrix as a result of this shift (see
Table~\ref{shifts}).
Information filtering later became known as \mbox{\em content-based}
filtering to the recommender system community and
has been applied to recommend movies~\cite{movie-content} and
books~\cite{bookContent}.
Content-based systems model content features of artifacts, rather
than of documents, and recommend items by querying such product features
against keywords or preferences supplied by the user~\cite{content-based}.
SDI (Selective Dissemination of Information), one
of the first information filtering systems, was based on keyword
matching~\cite{keyword}.  Content-based filtering is most
effective in text-intensive domains, which account for only a portion of
the artifact landscape.  Since we take a connection-oriented
perspective toward recommendation, content-based models and methods do not
find place in this survey.

In addition to identifying these differences, articles in this
special issue also reported new research developments.
\mbox{Foltz} and
\mbox{Dumais} introduced latent semantic indexing as a viable technique
to reduce dimensions in a \mbox{term-document} matrix~\cite{lsiCACM}.
More importantly for recommender systems,
\mbox{Goldberg}~et~al. coined the phrase
{\em collaborative filtering}~\cite{cf}
while describing \mbox{Tapestry}, which later became
known as the first recommender system~\cite{recSys}.
Collaborative filtering, which can be defined as
harnessing the activities of others in satisfying an
information-seeking goal, introduced another shift in IS
research.  Collaborative filtering
entails filtering items for a user that similar users
filtered.  Instead of computing artifact similarity (\mbox{content-based}
filtering), collaborative approaches entail computing user similarity.
The most salient difference between these two approaches is that
in content-based filtering users do not collaborate to improve
the system's model of them, while in collaborative approaches users
leverage the collective experience of other users to enrich
the system's model.  Collaborative filtering is
predicated upon persistent user models, such as profiles,
which encapsulate preferences and features
(e.g.,~married), rather than ephemeral queries.

This shift replaced features with representations of people (e.g., rating or
profiles) to filter documents in a modeling matrix.
While documents still constituted the other dimension of the
matrix, the word `document' assumed a broader meaning after the birth of the
web.  In addition to its traditional interpretation, it
also came to mean webpages and bookmarks~\cite{Fab,Siteseer,PHOAKS}, as well as
Usenet and e-mail messages~\cite{cf,GroupLens}.

Collaborative-filtering is effective since people's tastes are
typically not orthogonal.  However, initially it was not embraced.
Meanwhile, the advent of the web and its
widespread use, popularity, and acceptance,
made reducing information overload 
a necessity.  Of particular importance was social information filtering, a
concept developed by Shardanand and Maes~\cite{socialFiltering}.
A few years later, in 1996, interest in collaborative filtering led to a
workshop on the topic at the
\mbox{University}~of~\mbox{California},~\mbox{Berkeley}.
The results of this \mbox{Berkeley} workshop led to the
March 1997 {\em Communications of the ACM} special issue on {\em recommender
systems}, a phrase coined by \mbox{Resnick} and \mbox{Varian}
in their article introducing the issue~\cite{recSys}.

\begin{table}
\centering
\begin{tabular}{lr} \hline
{\bf Concept} & {\bf Modeling Matrix} \\
\hline
\hline
Information Retrieval & terms $\times$ documents \\ 
\hline
Information Filtering & features $\times$ documents \\
\hline
Content-based Filtering & features $\times$ artifacts \\
\hline
Collaborative Filtering & people $\times$ documents \\
\hline
Recommender Systems & people $\times$ artifacts \\
\hline
\end{tabular}
\caption{Shifts in matrix models outlining the evolution
of recommender systems from information retrieval.}
\label{shifts}
\end{table}

Resnick and Varian choose the phrase `recommender systems' rather than
`collaborative filtering' because recommenders need not
explicitly collaborate with recommendation recipients, if at all
(helping to reconcile the differences between content-based and collaborative
approaches)~\cite{recSys}.
Furthermore, recommendation refers to
suggesting interesting artifacts in addition to solely filtering
undesired objects (helping to reconcile the differences between IR and IF).
Resnick and Varian define a recommender as a system which
accepts user models as input, aggregates them, and returns recommendations to
users.
Two early collaborative-filtering recommender systems were
Firefly and LikeMinds.  Firefly evolved from Ringo~\cite{socialFiltering}
and HOMR (Helpful Online Music Recommendation Service) and allows a
website to make intelligent book, movie, or music recommendations.
Firefly's underlying algorithm~\cite{Shardanand}
is now used to power the recommendation
engines of sites such as BarnesandNoble.com.

Collaborative approaches constitute the main thrust of
current recommender systems research.  Once users are modeled, the
process of collaborative filtering can be viewed operationally as a function
which accepts a representation of users  and universal set of artifacts
as input and returns a recommended subset of those artifacts as output.
More importantly for this survey,
recommender systems also are intended to connect
groups of individuals with similar interests and to leverage
the collective experience rather than merely focusing on the
information-seeking goal of a specific individual~(as in a 
typical IR setting).
In order to make connections, this function typically computes
{\em similarity}~(e.g.,~closeness, distance, or nearest neighbor).
Making recommendations and thus connections
then entails approximating this function.
Approaches to this approximation that have evolved range from
statistical models~(e.g.,~correlating user ratings~\cite{GroupLens} or
reducing dimensions~\cite{eigentaste})
to attribute-value based learning techniques
(e.g.,~decision trees, neural networks, and
Bayesian classifiers)~\cite{AIMA} and have demonstrated
qualified success~\cite{CFalgos}.
Ultimately these techniques can be viewed as ways to infer structure
and induce connections in the modeling matrix space.

This final shift replaced documents with artifacts in the modeling matrix.
While the evolution of recommender systems research is characterized
by the shifts in matrix models illustrated in Table~\ref{shifts},
the sparsity and \mbox{anti-symmetric} properties 
remained constant across each. As shown below,
the web makes the matrix model symmetric.
Sparsity is mostly attributable to the reluctance of users to rate artifacts.
Reluctance results from a lack of time, patience, or willingness to
participate.
Sometimes the benefits gained from
providing constructive feedback are not apparent initially.
Reluctance may be partially attributable to a heightened awareness
of privacy when divulging personal information.
Therefore, collaborative-based recommender systems must
mediate an accuracy (of connection) vs. sparsity (of model) tradeoff.
The following two sections are devoted to strategies for filling in
cells of the initially sparse modeling matrix.

Since 1997 recommender systems research has advanced in many 
directions, such as reputation systems~\cite{repSys} (e.g.,~eBay.com),
and was placed
in a larger context called `personalization'~\cite{persViewsOfPers}.
The functional-emphasis of current recommender systems makes them
`templates for personalization'~\cite{piis}.

\section{Creating Connections: Explicit User Modeling}
\label{user-modeling}

\begin{table}
\centering
\begin{tabular}{ccccc}
\hline
& \multirow{7}{*}{\rotatebox{90}{({\em exploration} vs. {\em exploitation})}} & & & user reluctance to rate items (compounded by volume \& concern of
privacy)\\
& & & & $\downarrow$\\
\multirow{7}{*}{\rotatebox{90}{sustain}}
& & & & sparse modeling matrix (cold-start)\\
& & & & $\downarrow$\\

& & \multirow{3}{*}{\rotatebox{90}{$\longrightarrow$}} & $\longrightarrow$ & explicit + implicit user modeling ({\em exploration})\\
& & & & $\downarrow$\\
& & & & representation of user (ratings, profiles) as basis for connection\\
& & \multirow{1}{*}{\rotatebox{90}{$\longrightarrow$}} & & $\downarrow$\\
& & & $\longleftarrow$ & deliver recommendations \& create connections ({\em exploitation})\\
& & & & \\
\hline
\end{tabular}
\caption{User modeling methodology of a collaborative-filtering recommender
system.}
\label{methodology}
\end{table}

User modeling entails
developing representations of user needs, interests, and taste and is a
critical precursor to connecting people via recommendation.
In addition to personal characteristics,
users can be modeled by their assessments of products in the form of ratings,
which then become matrix entries.  Sparse user feedback is the single greatest
bottleneck of any collaborative-filtering algorithm:
`Collaborative filtering algorithms are not deemed universally acceptable
precisely because users are not willing to invest much time or effort in rating
the items.'~\cite{horting}.
These problems are compounded in voluminous domains, where a
large cumulative number of
ratings is required to sufficiently cover an entire set of items.
Moreover, as the number of dimensions (e.g., people or products)
grows larger, the number of multidimensional comparisons
grows.  In such situations
techniques from data warehousing and OLAP~(On-Line Analytical Processing)
are applicable~\cite{mutliDRS}.
In large domains, users typically examine and evaluate only a small
percentage of all items.
Shallow analysis of content makes fostering
connections difficult since opportunity for user overlap is limited.
While in the initial stages of a system, this challenge has been echoed as the
`cold-start' problem~\cite{cold-start}~(also referred to as the
`day-one' or `early-rater' problem),
it is also ubiquitous during the lifetime of a system.
For example, a collaborative
recommender has no platform to compute connections for a new
user who has yet to rate products or a new item which has yet
to be evaluated.  Such problems in developing a basis for collaboration
provide ample motivation for hybrid approaches which employ
content-based filtering in these specific
situations.  Hybrid systems have shown improved performance
over either single focus (pure) approach~\cite{joinCBFandCF,early-rater,hybrid}.
Systems must collect user data which affords the 
identification of differences, commonalities, and relationships among people.
In short, the goal is to add more and more information
to transform a sparse matrix to a dense matrix with added structure.

Approaches to user modeling can be studied by how they harvest
data~\cite{recSys}, either explicitly by asking users to submit feedback
through surveys~\cite{GroupLens} or inferring user interest implicit in
(usage) data~\cite{inferring-user-interest,PHOAKS}.
Strategies for the former approach
are showcased in this section, while those for the
later are discussed in Section~\ref{implicit}.
The most important tradeoff to consider in user modeling is minimizing user
effort while maximizing the expressiveness of the representation
(as well as privacy).  In other words, there should be a small learning
curve.
Explicit approaches allow the user to retain control over the
amount of personal information supplied to the system, but require an
investment in time and effort to yield connections.
Implicit approaches, on the other hand,
minimize effort, collect copious amounts of (sometimes noisy) data, and make the
social element to recommender systems salient,
but raise ethical issues.
The secretive nature of these approaches often make users feel as if they
are under a microscope.  The user-modeling methodology for a
collaborative-based system is illustrated in Fig.~\ref{methodology}.

In explicit user modeling, evaluations~\cite{GroupLens} and profiles~\cite{Fab}
are provided directly
by users to declare preferences in response to solicitations for data
such as surveys.
Evaluations of recommended artifacts can be both quantitative (e.g.,~ratings),
akin to relevance feedback in IR and IF~\cite{SIFTER},
and qualitative (e.g., lengthy reviews at Epinions.com).
They also can be positive or negative.
In a hand-crafted profile, a user states interests through items such as
lists of keywords, pre-defined categories, or descriptions.
The system then matches other users against this profile to
recommend incoming artifacts.
Systems which take such an approach to user modeling are
SIFT~\cite{SIFT} and Tapestry~\cite{cf}.

Without crossing over to an implicit approach,
researchers have identified strategies to deal with reluctance
to make an explicit feedback requirement less noticeable and
taxing~\cite{GroupLens,recSys}.
Possible approaches to motivate users to evaluate items are
subscription services, incentives, such as transaction-based compensations,
and exclusions~\cite{provisionRec}.  Employing a 
pay-per-use model for recommender systems, where human experts
rate items, is a viable, though less dynamic, option.
While this approach connects users through experts and is thus 
collaborative,
it deemphasizes the naturally social (and personal) element to recommenders.
Default votes are another way to deal with sparse ratings~\cite{CFalgos}.
Developing and tightly integrating
natural user interface (UI) mechanisms to solicit and capture
feedback with existing interfaces for recommendation delivery 
may lead to less intrusive interaction and thus more
cooperation and data~\cite{RSinterfaces}.  A similar approach is to build
recommendation into
everyday systems, such as e-mail, news, and web clients, and
services like collaborative
spam detectors~(e.g.,~Cloudmark's SpamNet, http://www.cloudmark.com).
In addition to helping to collect more
explicit ratings, building recommendation into common UIs
may help disseminate recommender systems to the
masses.  Requiring users to evaluate clusters of, rather than individual,
items is another approach to minimizing effort.
Rather than tackling sparsity from a user perspective
in an explicit approach, it also can be approached from a system
viewpoint.  Filter-bots which automatically examine and
rate all products may occupy
empty cells of a modeling matrix~\cite{GroupLens2}.

Lastly, a problem endemic to the subjective nature of
explicit modeling techniques is that
some users are more effusive in their ratings than others.
Effusivity in ratings refers to cases of
users who share similar preferences, but
rate products on completely different scales.
Identifying variations in
rating patterns is an approach to combat effusivity~\cite{horting,effusivity}

\subsubsection*{Other Considerations}

A variety of representations have been used to store
user data~\cite{machineLearningUserProfiles}.
The lack of standards to represent such
information and its sources~(e.g.,~logs)
in a uniform manner make interoperability among recommender systems
a challenge~\cite{multipleSources,broaderPers}.
Cookies are mechanisms for capturing and storing user preferences,
often employed in e-commerce~\cite{causticCookies}.
While cookies combat the stateless HTTP protocol, like many of these techniques,
they raise security and privacy concerns because they are
typically unknowingly enabled and as a result personal information is
divulged.

A challenge for any user modeling approach (explicit or implicit, for
content-based or collaborative recommendation) is the tradeoff between
{\em exploration} (modeling the user) and {\em exploitation} (using the
model to predict future ratings or make recommendations and connections), akin
to that in reinforcement learning~\cite{RI}.  Studying the connections which can be
made via recommendation and the resulting social network induced in a random
graph setting provides technical insight into this problem.
Mirza, Ramakrishnan, and Keller identify a `minimum rating constraint'
required to sustain a system and predict values for it based on
various experimental rating datasets~\cite{jumpingConnections}.

Ultimately the approaches to user modeling illustrated in this and the following
section are
used to connect people.
While a purely collaborative approach to recommendation is
widely accepted and employed, it is riddled with endemic problems.
User modeling must address more than just
sparsity.  For example, it is difficult to make connections to
users with unusual or highly specific tastes.  Furthermore,
connecting users with similar interests who have rated different items
(e.g.,~`we both read world politics online, but he ranked BBC.com
webpages, while
I ranked CNN.com pages') is challenging.  Over-specialization of
evaluated artifacts,
sometimes referred to as the `banana' problem~\cite{banana},
arises since frequently purchased items, such as bananas
in a grocery market basket, will always be recommended.
Conversely, some
products are seldomly bought more than a few times in a
lifetime (e.g.,~automobiles) and thus
suffer from a low number of evaluations.
Over-specialization which is grounded in the exploration vs. exploitation
dilemma can be addressed by occasionally
forcing exploration.  For instance, one can inject
randomness~(e.g.,~crossover
and mutation in a genetic algorithm or epsilon in a
reinforcement learning algorithm) into a model.
Recommended artifacts also can be partitioned into {\em hot} and
{\em cold} sets, where the latter is
intended to foster exploration and 
increase the (rating) coverage of items in the system~\cite{horting}.

\subsection{Review of Some Representative Projects}

The following collaborative-based systems employ many of the 
explicit user modeling
techniques showcased above and illustrate what can be achieved with
representations of users.   People are connected in the following systems
through statistical~\cite{eigentaste,GroupLens},
agent-oriented~\cite{Fab}, and 
graph-theoretic~\cite{horting} approaches.

\subsubsection*{GroupLens}

GroupLens recommends Usenet news messages~\cite{GroupLens}.
The system models users directly by explicitly eliciting
and collecting ratings of messages through an independent newsreader.
GroupLens is a project of the
recommender systems research group at the University
of Minnesota.  Usenet news is a
personal, voluminous, and ephemeral media~(in comparison to movies) and
thus an excellent candidate for collaborative filtering.
A total of 250 people
evaluated over 20,000 news articles~\cite{GroupLensOrg}.  GroupLens takes a
statistical approach to making connections.  The system predicts how
a user seeking recommendation would rate an unrated article by
computing a weighted average of the ratings of that message by users
whose ratings were correlated with the user seeking recommendation.
Correlation is computed with Pearson's $r$ coefficient.

A research issue is deciding whether to provide personalized predictions~(as
GroupLens currently does) vs. personalized averages.  Empirical research using
Pearson's $r$ correlation coefficient
revealed that `correlations between ratings and predictions is 
dramatically higher for personalized predictions than for all-user
average ratings'~\cite{GroupLens}. These results reinforce the hypothesis
that not all users are interested in the same
articles even within a certain newsgroup (e.g., consider a vegetarian
vs. a meat-eater in rec.food.recipes).

Konstan~et~al. state that `predictive utility is the
difference between potential benefit and risk'~\cite{GroupLens}.
The potential benefit of making predictions is the value of hits and
correct rejections.  The risk involved in making predictions is the cost of
misses and false positives.
Konstan~et~al. identify many of the approaches to increasing rating coverage
in explicit user modeling discussed above, such as
filter-bots---programs
that automatically read and rate all articles~\cite{GroupLens2}.
They also identify implicit declarations of quality such as the time spent
reading an article.  Konstan~et~al. recognize that
some users are more effusive with their ratings than others.
The developers of GroupLens began
NetPerceptions~(http://\hskip0ex www.\hskip0ex \mbox{netperceptions}.\hskip0ex
com), a company
employing collaborative filtering to provide personalization solutions.

\subsubsection*{Fab}

Balabanovi{\'{c}} and Shoham take an agent-oriented
approach to web document recommendation in Fab,
which grew out of a Stanford University digital library project~\cite{Fab}.
People are modeled in Fab through explicit (and some implicit) techniques
resulting in ratings and profiles.
The construction of accurate user profiles
drive various agents in dynamically adapting the system.
Fab is therefore a representative illustration of the importance and power of
user modeling.
The hybrid approach to recommendation in Fab retains the advantages of
both a content-based and collaborative approach while addressing the
disadvantages of each.  Moreover the
synergy yields new benefits.
Fab treats each single focus (pure) approach to recommendation
as a special case of its hybrid of the two.
`If the content analysis component
returns just a unique identifier rather than extracting any features, then
it reduces to pure collaborative recommendation; if there is only a single
user, it reduces to pure content-based recommendation'~\cite{Fab}.

Fab consists of two processes:
collection and selection.  During the collection phase, agents
gather web sources on topics discovered from clustering user profiles.
In the selection process, an agent matches a user profile against what the
collection agent has gathered.  Thus, one user may be matched against many
topics and multiple users may be interested in the same topic.  
Users proceed to rate results.  A user's ratings modifies his
profile and concomitantly
helps the collection agent harvest more relevant information
for an updated profile.  Fab uses

\begin{quote}
`the overlaps between
users' interests in more than just collaborative selection.  The design of
the adapting population of collection agents takes advantage of these
overlaps to dynamically converge on topics of interest \ldots and providing
the possibility of significant resource savings when increasing the
numbers of users and documents'~\cite{Fab}.
\end{quote}

\noindent
Fab connects two users if
a collection agent has clustered their profiles in order to collect more
web sources of the central theme of the cluster.

\subsubsection*{Intelligent Recommendation Algorithm}

A graph-theoretic collaborative filtering algorithm developed as
part of the suite of recommendation engines in the Intelligent Recommendation
Algorithm (IRA) program at IBM Research is presented in~\cite{horting}.
The algorithm is motivated by sparsity;
Aggarwal et al. contend that most collaborative filtering
algorithms, such as those for Firefly, LikeMinds, and GroupLens,
rely on too many ratings to be successful because they connect
users {\em directly} (e.g., users $A$ and $B$ are connected if their ratings
for at least $n$ items are correlated).  These algorithms compute
{\em closeness} by taking a weighted average of
{\em only immediate neighbors}.  Rather than
viewing sparsity as a vice, Aggarwal et al. exploit it in their algorithm.
The greatest contribution of their algorithm is its use of
functional {\em indirection}, i.e.,
it allows recommendations to propagate, via more than one intermediary,
from a user to another who has not rated common items.
The idea is to form and maintain a directed graph, where vertices represent
users and edges represent {\em predictability}.
Since the graph is directed, recommendations are anti-symmetric.
When one user predicts another, ratings propagate in this model.
`The ultimate idea is that predicted rating of item $j$ for user $i$
can be computed as weighted averages computed via a few reasonably short
directed paths joining multiple users'~\cite{horting}.  This effectively
makes predictability more general then closeness and addresses the
effusivity of ratings.

Aggarwal et al. also partition the presentation of
resulting recommendations into {\em hot} and
{\em cold} sets.  The hot set, which is two orders of magnitude smaller than
the cold, is intended to increase commonality to
provide better recommendations while the cold set is intended
to foster more exploration and increase the rating coverage of objects.
While most of the collaborative filtering systems with explicit
user modeling deliver predictable recommendations, this approach encourages
{\em creative} links which violate pre-existing
hierarchical classifications to introduce the possibility of {\em
serendipitous}
recommendations.
Aggarwal et al. evaluated their algorithm as well as that for
GroupLens, Firefly, and LikeMinds, for accuracy against synthetic data.

\section{Discovering Extant Social Networks}
\label{implicit}

Recognition of implicit declarations of user interest is a
precursor to discovering existing communities of people.  Implicit
modeling techniques
are a natural extension of those addressed in the previous section.
We begin by discussing projects which mine declarations of interest
in news~\cite{PHOAKS} and bookmark~\cite{Siteseer} datasets to cast
recommendations and make connections.  Although these systems foster
communities (as do all recommender systems),
they do not make social network identification salient.
While active effort is required when extracting declarations of interest
to make connections, natural connectivity among people is self-evident in
data.  We therefore next discuss mining connections implicit in 
communication~\cite{socialNet} and news~\cite{ReferralWeb} logs
to induce existing social networks to exploit for recommendation.
We then discuss mining and modeling social structure in movie ratings
datasets~\cite{jumpingConnections} and on the web~\cite{KleinbergAuthoritative}
via link analysis to help identify cyber-communities.
We conclude by discussing small-worlds~\cite{WSsmallWorld},
a new class of social networks which present compelling opportunities
for serendipitous recommendation.  Table~\ref{landscape} outlines the
landscape of research showcased in this section.

\begin{table}
\centering
\begin{tabular}{lclcl} \hline
{\bf Concept} & & {\bf Implicit declaration of interest} & & {\bf System}\\
\hline
\hline
Implicit User Modeling & & URLs in Usenet news & & PHOAKS\\
                       & & bookmarks & & Siteseer\\
                       & & & & \\
\hline
Link Analysis and Cyber-Communities & &  e-mail logs & &  Discovering Shared Interests\\
                        & & web documents & & Referral Web \\
                        & & & & \\
\hline
Mining and Exploiting Structure & & movie ratings datasets & & Jumping Connections\\
                                & & hits-buffs, half bow-tie & & \\
                        & & web link topology & & HITS\\
                        & & authorities and hubs & & CLEVER \\
                        & & bow-tie & & \\
                        & & & & \\
\hline
Small-World Networks & & actor collaborations & &  \\
                     & & author collaborations & & \\
                     & & infectious disease & & \\
                     & & the web & & \\
\hline
\end{tabular}
\caption{Recommender systems research focused on discovering existing
social networks.  The left column contains modeling concepts, while the
center column contains examples of implicit declarations of
interest or connections mined from the systems in the right column.
Notice that each system relies solely on structural, rather
than semantic, information.  Note also that the emtpy cell in the lower
right hand corner of this matrix
is a reflection that few systems take advantage of small-world
properties.}
\label{landscape}
\end{table}

\subsection{Implicit User Modeling}

Implicit approaches toward modeling users were developed in
response to the pervasive reluctance 
to evaluate recommended artifacts and, although less emphasized,
the possibility of building richer representations than with
explicit approaches.  The idea is to glean user preferences,
often secretively, by observation, to serve as surrogates for explicit
ratings.
A cold-start is less evident in this approach as
implicit ratings bootstrap the model and system.
Most of the techniques for implicitly
gathering and exploiting user information are based on 
methods and algorithms from machine
learning~\cite{machineLearningForUserModeling,Papatheodorou,Pazzani} and
data mining, which attempt
to discover interesting patterns or trends from large and diverse data sources.
These techniques are largely based on heuristics.
Data mining algorithms also have been subsequently
used to make recommendations.  Their expensive time and space complexity
is acceptable for user modeling which can be conducted offline.
Using data mining techniques for recommendation is a challenge because
recommendations must often be delivered in real time~\cite{Amazon}.
In addition, although not the focus here, inexpensive
and less complex techniques for computing recommendations
(e.g., correlating user ratings) are relatively effective.
Sophisticated approaches to recommendation also suffer from
yielding recommendations which are difficult to explain or believe;
recommendation explainability and believability are desired
properties~\cite{explainableRecs}~(see Section~\ref{broadening}).

A variety of data sources exist, teeming with and useful for gleaning
information about a user's interests and background.
Persistent keywords can be extracted from previous user queries.
Clickstream data, such as (web) access logs, are invaluable for
monitoring, capturing, and chartering
a user's interaction with a system, also called
`footprints'~\cite{footprints}~(e.g.,~actions during web browsing, links
followed, or amount of time spent on each product page).
Web log mining techniques~\cite{WebUsageMining} are therefore relevant to
this approach and have been used to create a platform to
recommend webpages based on browsing
similarities with previous users~\cite{autoPersWebMining}.
Web log mining also
has been used to trace patterns of navigation to restructure~\cite{adaptive}
and evaluate~\cite{evaluation} websites in a broader
personalization context~\cite{WebMining}.
Other techniques harness UI events
such as scrolling and mouse clicking~\cite{Goecks00}.
Alternate implicit indicators of preference include market baskets and
purchase transaction data, which are typically exploited by algorithms
for mining association rules~\cite{miningAssociationRules}.
Other, less obvious, implicit, self-organizing,
and social declarations of interest are bookmarks~\cite{Siteseer}.
The following two projects mine data sources containing implicit
declaration of interest for user modeling.  Based on their
mined implicit ratings they make recommendations and connections.
Although as a result of implicit user modeling virtual communities
are formed, these projects do not make identifying such communities
salient.

\subsubsection*{PHOAKS}

\begin{figure}
\centering
\begin{tabular}{c}
\includegraphics[width=7.25cm,height=3.25cm]{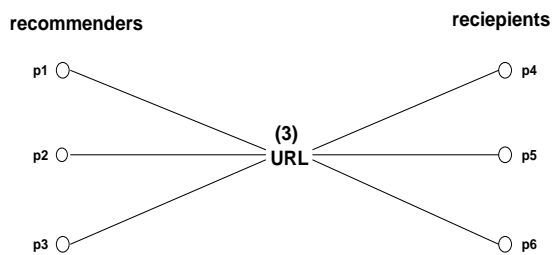}
\end{tabular}
\caption{PHOAKS' model of connecting users.
The three recommenders (p1, p2, and p3) have mentioned the
same URL in Usenet messages and thus that URL is recommended to the
recipients (p4, p5, and p6) with a weight of three.  This
illustration of the recommendation process of PHOAKS makes nine connections
among people (each recommender is connected to each recipient and
vice versa).}
\label{PHOAKS}
\end{figure}

Like GroupLens, PHOAKS~(People~Helping~One~Another~Know~Stuff)~\cite{PHOAKS}
recommends Usenet news messages, but unlike GroupLens, conducts
implicit, rather than explicit, user modeling.  PHOAKS interprets
the inclusion of uniform resource locators~(URLs) in messages
as an implicit declaration of interest.
The recommendation process in PHOAKS entails mining URLs,
filtering irrelevant and spurious URLs via heuristics~(e.g.,~remove links
embedded into an e-mail signature), and
computing a weight for each.  A link's weight is its occurrence
frequency in messages or, in other words, its number of distinct recommenders.
This metric was
later extended, formalized, and termed `authority weight' by
Kleinberg~\cite{KleinbergAuthoritative}.  Finally,
relevant recommended URLs and associated weights are returned to the user.
Precision and recall are applicable to the URLs that PHOAKS
recommends and thus were computed.
The recommendation process of PHOAKS,
which is illustrated in Fig.~\ref{PHOAKS}, connects a
recommender with a recipient if the recommender has included a URL
in a message in the recipient's search topic domain, and if that URL is
cited with a high frequency in that domain. 

Terveen~et~al. evaluated their weight metric by
comparing PHOAKS recommendations
to those provided in frequently asked question lists~(FAQs) which
are created by human judges of
quality. Their evaluation approach suggests a novel application of PHOAKS; it
also can be used to
semi-automatically create FAQs or recommend improvements to extant FAQs.
In addition, Hill and Terveen present an idea for improving
the quality of search engine results called community sorted
search~\cite{communitySortedSearch}.  The idea is to
run a keyword search and cluster
the results based on the newsgroups which mention each link in order
to disambiguate the query.  URLs could be presented sorted by
frequency within messages of each cluster.

\subsubsection*{Siteseer}

Siteseer models users through their bookmark folders~\cite{Siteseer}.
Bookmarks are a rich data source to exploit for user modeling
because they obviate the need to collect explicit ratings and
are an implicit declaration of interest,
self-maintained, and less noisy in comparison to
other implicit rating sources such as
a mouse click~(which could be random) or a URL embedded in a newsgroup
message~(e.g.,~`[URL] is uninformative!')~\cite{PHOAKS}.
Furthermore, the binary nature of a bookmark~(presence or absence)
eliminates the possibility of partial preference.  Bookmarks do not
capture lack of preference as do other data sources.
Most importantly
bookmark folders are the basis for the formation of a virtual community.
Siteseer computes set intersection between input
bookmark folders.  A user who has a folder with
the greatest overlap with the seeking user's folder is the best
qualified recommender for that seeker in the context of that folder.
Furthermore each individual URL can be
assigned more weight based on the number of folders that it appears
in within a virtual community, akin to the authority weight metric
of PHOAKS. Siteseer recommends a set of bookmarks in context~(i.e.,~a folder).
The system connects people in a directed, non-reciprocal manner
akin to IRA~\cite{horting}.  It is pertinent to note that
Siteseer does not work on any semantic information such as bookmark title
and is therefore an illustration of how indicative purely~(social) structural
information can be.  Siteseer suffers from problems endemic to a purely
collaborative approach.  For instance,
both a new user to the system and an existing user
who want to create a new folder provide Siteseer no input basis to
compute overlap.  Conversely, no collective experience is available
to leverage until a cyber-community has been discovered.
In addition, bookmarks are typically not public domain
or readily available.  Siteseer therefore requires trust and buy-in.

\subsection{Link Analysis and Cyber-Communities}

Siteseer is on the verge of directly identifying communities implicit in
data.  A natural extension of Siteseer is to proactively
mine data which saliently reveals social connections among people.
In addition to explicit communities, such as
discussion lists, \mbox{e-groups}, and community portals~\cite{BEV},
many communities also are implicit
in data, such as communications logs~\cite{socialNet} and
webpages~\cite{ReferralWeb}, which are fertile reflections of natural
connectivity among people.  These communities are available to be
identified, explored, and exploited~\cite{selfOrg}.
Identification is also worthwhile since, unlike connections induced via
explicit modeling approaches and the corresponding systems in the
previous section, implicit social networks foster the attractive
possibility of {\em serendipitous} collaborators and recommendation.
For example, consider
the editor of a journal interested in forming an impartial committee of 
reviewers for a submitted paper.  A social model of author
collaborations is an invaluable resource for such a task.  Furthermore,
unlike other recommender
systems which require users to create and maintain profiles~\cite{Fab},
approaches which model people connections or social organization
implicit in rich,
self-generating data result in representations which are
likely to be more accurate reflections than a user's perception of
his own connections~\cite{ReferralWeb}.
These communities are typically modeled as social networks~\cite{SNA} and
thus research from the social network analysis community is relevant to
(automatically) discovering and exploring these networks.
In this section we emphasize automatic social network
induction, exploration, and exploitation, especially since the
manual identification and formation of such communities, collaborators, and
referral chains is painstaking, error-prone, and time-consuming.

\subsubsection*{Social Networks}

\begin{figure}
\centering
\begin{tabular}{c}
\includegraphics[width=5.75cm,height=4.5cm]{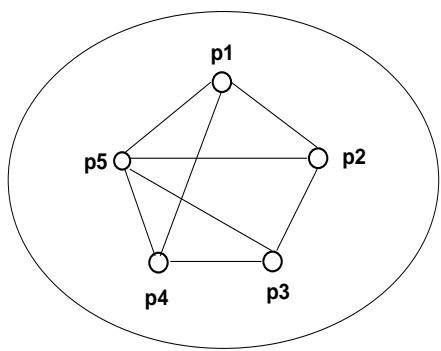}
\end{tabular}
\caption{A friendship social network graph. Vertices represent people
and an edge between two vertices denotes the `friend-of' relation.} 
\label{SocialNetwork}
\end{figure}

The study of social network analysis dates back to the 1960s.
Social networks, which derive their name from social associations among people,
model a social process, or specifically,
connections among individuals or objects.
A social network graph is a unipartite undirected graph,
where vertices represent objects (typically people) and edges
represent relationships between those objects (e.g., `friend-of').
A social network graph is thus characterized by heterogeneous vertices and
homogeneous edges, i.e.,~while each vertex represents a unique 
entity, each edge represents the same relationship.
Fig.~\ref{SocialNetwork} illustrates a friendship social network reproduced
and enlarged from the center of Fig.~\ref{overview}.

Although not the focus of this paper, it is relevant to note that
many interactions and                                                            associations in existing web information spaces 
are modeled via a social network navigation 
metaphor~\cite{ReferralWeb,compNetAsSNs}.
Two popular examples are the Internet Movie Database at
http://\hskip0ex www.\hskip0ex imdb.\hskip0ex com and the DBLP
bibliography website at
http://\hskip0ex www.\hskip0ex informatik.\hskip0ex uni-trier.\hskip0ex
de/\hskip0ex $\sim$ley/\hskip0ex db/, a collaboration graph
of authors, papers, journals, and conferences.
In contrast to hierarchical classifications, where users
systematically `drill-down' to hone in on desired information, in sites
based on social
network navigation users interactively `jump connections' in support of an
exploratory information-seeking goal.
Support for foraging in a network, modeling
connections among entities, may have been born out of
Vannevar~Bush's 1945 essay {\em As~We~May~Think} which
foreshadows the web~\cite{AsWeMayThink}.  In this
article Bush states that selection by associations among items more closely
matches human perception of information foraging than selection
by hierarchical index structures.

The increased interest in the concept of social networks has led to the
formation of 
communities and journals devoted to the subject.
The International Network for Social Network Analysis (INSNA), which
was founded by
Wellman in 1978 (http://\hskip0ex www.\hskip0ex heinz.\hskip0ex
cmu.\hskip0ex edu/\hskip0ex project/\hskip0ex INSNA),
has emerged as an authority on social network analysis.
Journals on the topic include {\em Social Networks},
{\em Connections}, and the {\em Journal of Social Structure}.  Note
however, that the main thrust
in the above forums remains targeted toward the sociological,
rather than the computing, community.  Studying
algorithms which identify and
exploit the combinatorial structure of social networks
is a newly emerging computer science research area being actively
investigated by Kleinberg at Cornell University.
We refer the interested reader
to~\cite{WebStrucScience,compNetAsSNs} for a succinct
introduction.

Following are examples of projects which mine connections and
exploit the resulting social network discovered for various
recommendation purposes.
The research projects showcased here attempt to identify social
networks in data, via various heuristics,
implicit in naturally generated data,
rather than mining user preferences to form them; they 
also emphasize modeling connections which
distinguishes them from those in the previous subsection.
The following two projects emphasis social network
induction, exploration, and exploitation.

\subsubsection*{Discovering Shared Interests}

One of the earliest yet untouted attempts at inducing
social networks by link analysis entailed analyzing
e-mail communication logs, based on
heuristics, to uncover an extant social network~\cite{socialNet}.
In the identified communication
network, two individuals share an edge if an e-mail was exchanged
between the two.  Spurious connections were pruned by heuristics.
The discovered social network was intended to
help people (vertices) identify others with similar interests and
encourage cross-fertilization among the cluster participants.
The authors defined
the concept of {\em closeness} between two vertices with
the following function:

\begin{displaymath}
InterestDistance(n_{1},n_{2}) =
\frac{|(C(n_{1}) \cup C(n_{2})) - (C(n_{1}) \cap C(n_{2}))|}{|(C(n_{1})
\cup C(n_{2}))|}\textrm{,}
\end{displaymath}

\noindent
where $C(n_{i})$ is the set of vertices directly connected to vertex $n_{i}$.
A value of zero for this interest distance metric indicates that the two input
vertices have all neighbors in common. Conversely, a value of one indicates
that the two input vertices have no neighbors in common.
Such a function is useful for
information-seeking activities such as locating a known expert
on a particular subject.  The authors
contend that other individuals within close proximity of that
expert vertex are also recommenders on the particular subject.
This research was not embraced  when published in 1993.  However,
ten years later, copious amounts of log data as a result of the
ubiquitous nature and extensive use of the web has made this work attractive.
It is now frequently cited
in both the recommender systems and social network analysis literature.

\subsubsection*{The Hidden Web: Referral Web}

Akin to the work described
in~\cite{socialNet}, the Referral Web project at AT\&T Labs
also implicitly models users to form social networks, where
an edge exists between two individuals
if their names appear in close proximity in a web document~\cite{ReferralWeb}.
Again, the underlying assumption is that clustered
vertices correspond to people who share similar interests.
The Referral Web project also builds on
many of the ideas first introduced in~\cite{socialNet} by, e.g.,
exploring the resulting social networks to
find experts or recommenders on a particular subject.
Kautz et al. intended for users to interactively explore the implicit,
existing social network in web documents that their mining made explicit.
The authors discuss three types of information-seeking goals users
could attempt to fulfill in the resulting
network---finding referral chains, searching for experts, and
proximity search near known experts---which are reflected
in the context of computer
science researchers in the following information-seeking questions,
respectively~\cite{ReferralWeb}:

\begin{itemize}
   \item What is my relationship to Marvin Minsky?
   \item What colleagues of mine, or colleagues of colleagues of mine,
                know about simulated annealing?
   \item List documents on the topic `annealing' by people close
         to Scott Kirkpatrick.
\end{itemize}

Kautz et al. also have developed the complementary concepts of
{\em accuracy} and {\em responsiveness} in a social network~\cite{HiddenWeb}.
They hypothesized that the accuracy of a referral is inversely proportional
to the number of
intermediate links between the individual seeking recommendation and the
expert providing the recommendation (referred to as
`degrees of separation'---the length of the shortest path connecting
two vertices in a social network graph).
Kautz et al. refer to this in the model they developed as
the {\em referral factor} $A$, a real number between 0 and 1.
$p(A,d) = A^{\alpha d}$,
where $\alpha$ is a fixed scaling factor and $d$ is the number of steps from
the vertex to an expert. $A$ is the probability that a vertex will refer in the
direction of an expert.
Likewise, the further removed an expert is from a requester, the less
responsive the expert is expected to be.
Kautz et al. refer to this in the model they developed as
the {\em responsiveness factor} $R$,
again a real number between 0 and 1.  $p(R,d) = R^{\beta d}$,
where $\beta$ is a fixed scaling factor representing
the probability of an expert responding to a requester $d$ links away.
Kautz et al. ran a series of simulations on this model 
with various values for each of these parameters.  The results of the
experiments reveal a tradeoff between $A$ and $R$~\cite{HiddenWeb}.
Through simulation, Kautz et al. discovered that automatic methods
can outperform manual referral chaining.  Automatic
referral chaining however requires sending more messages and thus demands
the search of more vertices.  Identifying the
effects of certain parameters of a developed model for
social network graphs is invaluable 
for setting such parameters when
designing a system.  Such experimental analysis has been conducted on movie
rating datasets~\cite{jumpingConnections}.
An online demo of Referral Web is available
at http://\hskip0ex weblab.\hskip0ex research.\hskip0ex att.\hskip0ex
com/\hskip0ex refweb/\hskip0ex working/\hskip0ex temp/\hskip0ex
RefWeb.\hskip0ex html.

\subsection{Mining and Exploiting Structure}

Mining social networks from existing data
is a method of implicit user modeling for collaborative recommendation.
Social networks also can be formed by applying transformations
on other, typically bipartite, graph representations identified in datasets.
Entertaining the
possibility of inducing social network graphs from bipartite graphs fosters
social network analysis in
domains where the presence of such networks is not salient.
Consider that
a ratings dataset can be modeled as a bipartite graph rather than a matrix.
In social network theory, a
bipartite graph is referred to as an
{\em affiliation network}~\cite{SNA} (other researchers
refer to them as {\em collaboration graphs}~\cite{graphTheory2}).
In social network theory a {\em mode} is defined to be a
`distinct set of entities on which structural variables are
measured'~\cite{SNA}.
A social network graph consists of only one mode (e.g., `people' in
Fig.~\ref{SocialNetwork}) containing vertices which
share a unifying feature, while an affiliation network
has two modes---a {\em primary} mode and a {\em secondary}---each corresponding
to a disjoint vertex set in the bipartite graph.

A classical example of an affiliation network is the actor-movie collaboration
graph (also known as the `Hollywood graph'~\cite{graphTheory1}),
where actor and movie are the primary and secondary modes,
respectively.  In order to induce a social network graph,
members of the primary mode can be brought together
via their relationships with members of the secondary mode.
The role of the secondary mode is then to bring objects of
the primary mode together.  This process can be viewed as collaborative
filtering.  For example, consider that
the community of authors of computer science publications
is implicit in a computer science corpus or digital library.
This social network, where vertices represent authors and two authors share an
edge if they have coauthored a paper, can be mined from an
author-paper affiliation network representation of the corpus.
The Institute for Scientific Information (ISI) maintains such networks and
provides associated products and services to
help facilitate the discovery process for researchers.
An editor of a technical journal may
wish to employ a recommender system which models this network to facilitate
her formation of an impartial committee of reviewers to evaluate a paper
submitted by a particular modeled author.
In such cases, candidates would be those
with moderate closeness to the reviewee.  While an
impartial reviewer is one who has never coauthored a paper with the
reviewee, a prime candidate should also
have moderate knowledge of the submitted paper's topic and thus be
slightly close to the reviewee in the resulting social network.
In such cases, the editor may want individuals within two or three degrees
of separation from the author whose paper is to be reviewed.
When applied to such activities, recommender
systems which model such social connections among individuals are
compelling and applicable to a variety of other important information-oriented
tasks~(including some studied in the field
of `bibliometrics'~\cite{bibliometrics})
such as granting tenure to a faculty member of a university.

Inducing or identifying social networks
entails mining social structure in representative datasets which
typically results in a graph model.
That graph then can be explored and exploited
to support many information-seeking and recommendation-oriented activities.
Mining structure typically entails identifying
characteristics, such as the degree of connectivity or clustering
to make statements with certainty
about the underlying domain and social process.
Investigating why structure arises in the first place also is useful for
gathering insight into the underlying social process and its implications
on recommendation.
Exploring and exploiting graph structures of social processes is a viable and
increasingly popular way to satisfy information-seeking goals as
reflected in~\cite{SIGIR2003keynote,WebGraph,WebStrucScience}.
The following two projects entail explicitly identifying structure, such as
connectivity and level of clustering, and in turn exploit it for
recommendation.  The main idea of this
section is to mine, model, and exploit social structure for recommendation.

\subsubsection*{Jumping Connections}

\begin{figure}
\centering
\begin{tabular}{cc}
& \mbox{\psfig{figure=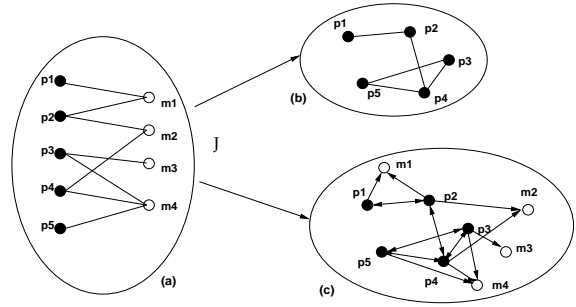,width=7.5cm,height=4.0cm}}
\end{tabular}
\caption{Inducing a social network graph (b) from an affiliation graph
representation (a) of a movie rating dataset via a skip jump.
Re-attaching movie vertices to
the social network graph yields a recommender graph (c) which forms
a half bow-tie structure.
Figure used from~\cite{BatulThesis} with permission.}
\label{skip}
\end{figure}

Mirza, Keller, and Ramakrishnan
developed a graph-theoretic model to design and evaluate recommender
systems~\cite{jumpingConnections}.  Their approach is connection-centric
and entails inducing social networks and identifying various
structural properties therein from public domain movie rating datasets.
Each ratings dataset $\mathcal{R}$ used, namely EachMovie (collected by the
Digital Equipment Corporation) and MovieLens (developed by the
recommender systems researcher group at
University of Minnesota for the MovieLens project which is based on the ideas
from GroupLens; see http://movielens.umn.edu),
was a matrix (people $\times$ movies)
of ratings and was modeled as an undirected bipartite graph,
where the two disjoint vertex sets correspond to people ($P$)
and the movies ($M$) they
have rated (Fig.~\ref{skip}a).
An edge between a vertex of each set denotes the
`rated' relationship.  In these affiliation networks,
people and movies are the primary and secondary modes, respectively.

Mirza et al. induced various social networks by applying various
`jump' specifications of how individuals in
the affiliation network representation of $\mathcal{R}$
are to be connected in the resulting social network graph $G_{s}$
(Fig.~\ref{skip}b).
A jump is a function $\mathcal{J}: \mathcal{R} \mapsto S$,
where $S \subseteq P \times P$.  The elements of $S$~(unordered
people pairs) represent the edges of $G_{s}$.
There are many ways of jumping.  A {\em skip jump}
directly connects person $x$ and person $y$ in $G_{s}$ if they have rated
a common movie.  The skip jump is a special case of a {\em hammock jump}, which
induces a $G_{s}$ where two people are
directly connected if they have evaluated $w$ movies in common, where
$w$ is termed the {\em hammock width}.
A jump can be thought of as a recommendation algorithm and 
therefore, all recommender systems bring people together,
in one form or another, via jumps,
albeit each system may jump in strikingly different ways.
Systems then can be classified by how they jump and the
number of people each jump
connects in the resulting $G_{s}$.  Mirza et al. also propose the number of
people connected by a jump as an evaluation metric for recommender systems
(see Section~\ref{broadening}).
Finally, re-attaching the movie vertices
to the people vertices of $G_{s}$~(Fig.~\ref{skip}b)
produces a recommender graph $G_{r}$ (Fig.~\ref{skip}c), where
directed paths from people to movies may be discovered.

After applying a jump,
Mirza et al. attempt to identify structural properties in $G_{s}$ and $G_{r}$.
They have identified a half bow-tie structure in $G_{r}$.
This higher-order structural observation is analogous to the identification
of the (full) bow-tie structure in the link topology of the web~\cite{bow-tie},
which arises since the web's nucleus consists of a strongly connected
component~(SCC, the center of the bow-tie), webpages which
link only into it~(the left side of the bow-tie), and pages which
are only linked to from the SCC~(the right side of the bow-tie).
In $G_{r}$, the $G_{s}$ is the SCC~(the center of the half
bow-tie) while the movie vertices of $G_{r}$,
the right half of the bow-tie, are only linked from this SCC.
Fig.~\ref{skip} illustrates the entire process by which a $G_{s}$
is induced from an affiliation network representing $\mathcal{R}$
via a skip jump and subsequently
augmented to $G_{r}$ by reattached vertices of the secondary mode
of $\mathcal{R}$~(movies) to $G_{s}$.
Note that people are brought together via movies
analogous to Kleinberg's authorities being brought together via hubs in the
affiliation topology of the web~\cite{KleinbergAuthoritative}.

While both data sets studied by Mirza et al.
are extremely sparse~($\ge$93\%), both are connected.
Another interesting discovery was that each dataset exhibited
a {\em hits-buff} structure,~i.e.,~some people~(buffs)
rate all movies while some movies~(hits) are rated by all people.
This intriguing structural property
is attributable to both the underlying domain 
and the hypothesized power-law distribution~\cite{power-law}
followed by movie ratings (and
other self-organizing systems, e.g., the web and airports).  Notice again that
this structural discovery resembles the mutually reinforcing relationship of
Kleinberg's authorities and hubs.

The power law distribution of movie ratings has other implications.
For example, as $w$ increases, $G_{s}$
becomes more disconnected.  
A hammock width of seventeen or greater disconnects $G_{s}$
in the MovieLens dataset.
What is surprisingly interesting however is that at
that breaking point the graph has one large SCC plus many isolated vertices
rather than many 
small SCCs.  Mirza et al. refer to this process as `shattering' the
graph.  This phenomenon is again attributable to the power-law
distribution in movie ratings.  One large SCC and many small
SCCs (i.e., communities) do not emerge when $G_{s}$ breaks
because movie viewers typically do not have strong biases in their tastes.
Mirza et al. contend that in other domains, such as books and music, small
communities representing the many diverse genres (e.g., 
country or jazz) in such domains will arise as $w$ is increased.
Thus, the group hypothesizes that books and music ratings do not
follow a power-law distribution.
Such hypotheses have yet to be
verified due to the inability to obtain datasets in such domains, and
thus leave scope for future work.  

Akin to the tradeoff between the referral and responsiveness factor in
Referral Web~\cite{HiddenWeb}, 
Mirza et al. report the effect of the hammock width $w$ on the
minimum rating constraint $\kappa$~(i.e.,~the minimum number of 
items users must rate prior to receiving recommendations) and
the various shortest paths used to route recommendations,
such as $l_{pp}$ (shortest path length in $G_{s}$),
$l_{r}$ (shortest path length in $G_{r}$), and $l_{pm}$ (shortest path length
from people to movies in $G_{r}$).
Identifying an optimal number of evaluations required to sustain a system
is invaluable from a design perspective.
Random graph theory plays a large role in this analysis since movie
ratings do not correspond to a particular graph, but rather a family of graphs.
Thus, while graph algorithms, such as
Dijkstra's single source and Floyd's all pairs
shortest path algorithms, can be applied, they will
not reveal any interesting properties in $\mathcal{R}$.
The researchers thus study random graph models for
recommender dataset graphs and associated attributes
such as degree distributions.
Random graph models typically accept
an edge probability and number of vertices as input and output a random
graph meeting such properties and constraints.

The ultimate goal of the Jumping Connections project is to develop a
model wherein the implications of certain parameters
(e.g.,~$w$ or $\kappa$)
on the structural properties of $G_{s}$ and $G_{r}$ guide
the design of a recommender system. 
Development of such a model entails studying the effects
certain jumps have on the properties of the resulting $G_{s}$.
Based on random graph models the researchers would like to say that 
$G_{s}$ is connected if some condition holds~(e.g.,~ $\kappa >$ 20).
Being able to make such statements with high probability is sufficient
from a theoretical computer science point of view.
Such information is critical not only to comparing
various recommenders systems but also
to providing designers answers to questions
such as `If I know that I only need a hammock width of four
to effectively make
recommendations, to what should I set the minimum rating constraint of our
new recommender?'  

Future work includes incorporating user ratings.  Thus far the
group has only been
considering the binary nature of ratings (presence or absence)
to make connections.
This again reflects that substantial insight into recommendation can be
achieved with purely structural
information, as in Referral Web~\cite{ReferralWeb} and Siteseer~\cite{Siteseer},
rather than semantic information such as ratings.
Mirza~et~al. also would like to formally test their
hypothesis that jumps (i.e.,~recommendation algorithms) never bring disconnected
components of the recommender dataset graph together.
The researchers hypothesize that
in certain pathological cases, such as when the dataset graph contains two
isomorphic subgraphs, the singular value decomposition~\cite{decomSum}
(also known as latent semantic indexing to IR
researchers \cite{LSI}) will connect disconnected portions of $\mathcal{R}$.
Jumping Connections is a novel research project; it helps
make a science out
of recommender systems which have traditionally lacked any sophisticated model
to design and evaluate systems.

\subsubsection*{HITS}

In addition to mining bookmark, communication, and news datasets,
recently much
research has been conducted on mining the link topology of the
web~\cite{miningLinkStr}.
The most significant and compelling contribution in this area is
Kleinberg's observation, through mining link structure, that the
web is an affiliation network consisting of an {\em authority} mode and a
{\em hub} mode.
Authorities are authoritative sources on a topic (e.g., PGA.com for golf)
while hubs are collections of links to authorities (e.g.,~bookmarks or
favorite links pages).  Hubs and authorities
mutually reinforce each other: good authorities are linked to
by many good hubs, while good hubs link to many good
authorities~\cite{KleinbergAuthoritative}.

The identification of this
compelling structural property was important since the web was
widely believed to be a unipartite
graph where all pages were perceived to be of the same type.
This observation gave
birth to search engines which rely exclusively on
purely structural information to navigate and search the web.
The identification of a second mode of webpages,
namely hubs, was required in order to connect authorities who would otherwise
not be linked due to competitive reasons~(e.g.,~the webpage of
\mbox{Microsoft} does not link to the webpage of IBM, yet both are computer
companies).  Kleinberg's  HITS (Hyperlink-Induced~Topic~Search)
algorithm exploits such structural information to
search the web for authoritative sources.

The HITS algorithm attempts to identify good hubs and authorities by assigning
hub and authority weights to webpages based on a (QR~\cite{decomSum})
matrix power iteration.
A similar project~\cite{trawling} approaches trawling the
web for cyber-communities as mining structure in bipartite graphs.
These researchers however describe Kleinberg's hubs and authorities
as fans and centers, respectively.  The two groups have collaborated on 
a survey of the measurements, models, and methods of the `web
graph'~\cite{WebGraph}.
Amento, Terveen, and Hill
experimentally measured whether authoritative sources
are good predictors of quality~\cite{authorityQuality}.

As a result of mining the link topology of the web,
the HITS algorithm is the most illustrative and powerful
example of what can be done purely with structural information akin to
Referral Web~\cite{ReferralWeb}, Siteseer~\cite{Siteseer}, and
Jumping Connections~\cite{jumpingConnections},
rather than semantic information, such as text indexed on a page.
Specifically, HITS is evidence that
link structure is sufficient to correctly characterize, aggregate, and
leverage the interests of a large population.

The CLEVER search engine of IBM~\cite{miningLinkStr}
was designed based on HITS.
Google, developed at Stanford University,
is another search engine which considers link structure~\cite{anatomy}.
PageRank, the algorithm of Google, only computes authority weights, however,
and thus does not connect authorities via hubs.
The Google
engine also analyzes textual information in addition to link structure.
It is therefore involves a hybrid approach (i.e., both
structural and semantic information is incorporated).
The most salient difference between the PageRank and HITS algorithms
is that PageRank analyzes the link structure of the web offline while CLEVER
mines the web on a per query basis.  An implication of this
difference is that Google is a much more practical application than CLEVER
due to the expensive matrix operations required for HITS in real-time.

\subsection{Small-World Networks}

While typical examples of social networks are the actor collaboration graph
and author collaboration graphs, a new class of social networks~(and
associated
random graph models) has emerged---{\em small-world
networks}~\cite{graphTheory2}.
These networks naturally model the small-world phenomenon.
In the late 1960's, Harvard social psychologist Stanley Milgram
paved the way for small-world network
analysis by conducting a unique chain letter experiment~\cite{Milgram}.
As opposed to the other projects discussed in this section,
rather than attempting to discover a social network, Milgram 
hypothesized that a social network existed and tested both if that
network exhibited small-world properties
and, more importantly, if individuals,
with only local views of the network, could successfully
construct short chains between members.

The experiment involved source individuals in
\mbox{Nebraska} or \mbox{Kansas} delivering a letter to a target person
in \mbox{Boston}, MA via intermediaries.
Source individuals were given only a few cursory biographical
characteristics of the target and
permitted to forward the letter only through individuals to which the source
was on a first name basis and so on.
As letters propagated east across the US, the
experiment revealed that any two individuals in the
acquaintance network of the United States
could be connected through a few intermediaries; or more formally
that the acquaintanceship network of the US
exhibited \mbox{small-world} properties.
Milgram's experiments specifically revealed that any two randomly picked
individuals residing
in the US were connected by no more than six intermediate acquaintances.
The small-world phenomenon later became popularly
known as `six degrees of separation,' after which
both a play and its movie adaptation have been named.
The degrees of separation between two vertices in a social network graph is
the length of the shortest path connecting the two vertices.
The problem of reducing the number of intermediaries in a social network
has historically been referred to as {\em information routing}.

When put into context the small-world phenomenon
does not seem outlandish.  Consider that a college student in
a large state university might be connected to the
president of the United States 
through five intermediaries on a first name basis
(e.g.,~student--professor--department chair--dean--university president--president of the US)
in a connection path of length six.
The impact of the small-world phenomenon is ominous when
studied from the perspective of infectious disease.

\begin{figure}
\centering
\begin{tabular}{cc}
& \mbox{\psfig{figure=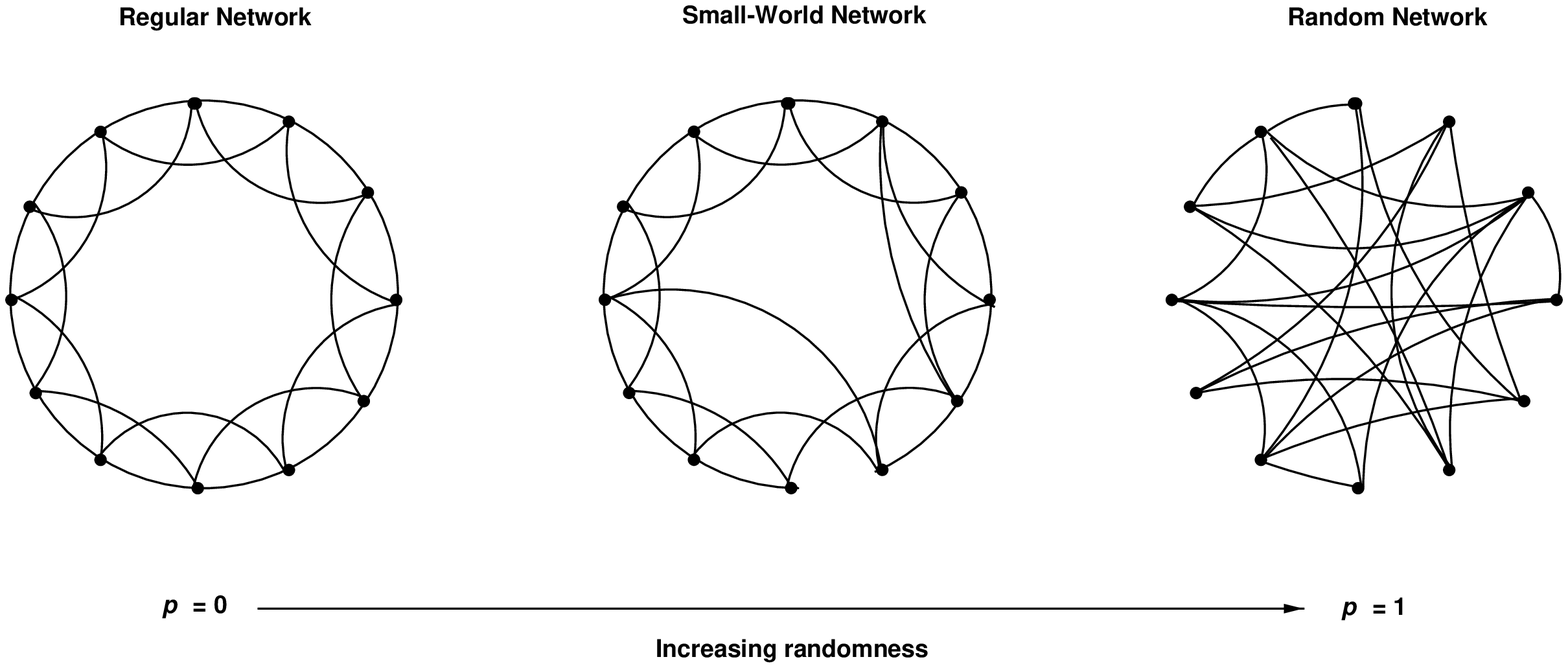,width=5in}}
\end{tabular}
\caption{The \mbox{Watts-Strogatz} model for \mbox{small-world} graphs.
Edges from a
regular ring lattice~(left) are randomly rewired based on a rewiring
probability $p$, yielding a \mbox{small-world} network after only a
few random rewirings~(center).  When $p$=1, all edges are rewired yielding
a completely random graph~(right).  Figure used from~\cite{BatulThesis} with
permission.}
\label{WSmodel}
\end{figure}

A small-world is a graph $G$ which exhibits certain structural
properties while modeling some natural phenomenon.
Small-world graphs are structurally
characterized by sparse edges (i.e., many more vertices than edges),
highly clustered vertices, and relatively short
paths between any two vertices.
Random graph models have been proposed to study small-world
networks~\cite{diameter,classesSmallWorld,Barabasi,Erdos}.
\mbox{Watts} and \mbox{Strogatz} have developed
a random graph model for \mbox{small-world} networks based on a rewiring
probability~(see Fig.~\ref{WSmodel})~\cite{WSsmallWorld}.
The \mbox{Watts-Strogatz} model
for \mbox{small-world} graphs interprets a \mbox{small-world} network
as a hybrid between a completely
ordered ring lattice (wreath)
and a completely random graph, leaning closer to the lattice.
The {\em average minimum path length} $A$  of $G$ is `the minimum
path length $L$ averaged over all pairs of vertices'~\cite{graphTheory2}.
Watts and Strogatz refer to average minimum path length as
`characteristic path length'~\cite{WSsmallWorld}.
The {\em clustering coefficient} $c$ of a vertex in $G$
is the degree of the vertex divided by
the  maximal number of edges which could possibly exist between the vertex
and all its neighbors.
Characteristic path length is a global property of a network, while 
clustering coefficient is a local property measuring the `cliquishness' in
a neighborhood.
Watts and Strogatz discovered that by replacing
some local contacts with arbitrary ones (called `random rewiring'),
the clustering coefficient
of a network remains high, close to that of the initial
completely ordered network, while the characteristic path length is drastically
reduced.  The model begins with a regular ring
lattice where vertices are highly clustered with large average minimum
path lengths
between any two vertices~(Fig.~\ref{WSmodel}---left).
In the model, from this lattice edges are randomly rewired based
on a rewiring probability $p$.  Randomly rewiring only a few edges (less than
ten percent of total edges) renders a
\mbox{small-world} graph where vertex clustering remains
relatively high, but average minimum
path length is drastically reduced in comparison to the ring
lattice~(Fig.~\ref{WSmodel}---center).  When all edges are rewired
the model produces a
completely random graph where vertex clustering is low and average path
length is small~(Fig.~\ref{WSmodel}---right).
The replacement of local links with random links (random rewiring)
manifests itself in the real world when someone moves to a new city,
starts a new job, or joins a club.

Watts and Strogatz hypothesized that small-world
properties exist in diverse domains and that
the small-world phenomenon arises in many
self-organizing systems such as
the actor-collaboration graph, the power grid of the Western
US, and the nervous system of the nematode worm Caenorhabditis
elgans.  They examined these real world networks
to test the existence of small-world properties.
All three observed networks exhibited small-world properties.
These conclusions suggest that the small-world phenomenon is common
in many large networks found in nature and not merely an artifact of an
idealized world.  Small-worlds also are believed to exist in
the spread of disease, and most importantly as of recently, the
web~\cite{graphTheory2,WSsmallWorld}.

The research of Watts and Strogatz has peaked interest in studying
many other diverse social domains via small-world graphs.
By conducting an exhaustive survey of the actor-collaboration graph
implicit in the Internet Movie Database,
Tjaden at the University of Virginia found the
maximum degrees of separation from any actor in
Hollywood to American film actor Kevin Bacon, termed the `Bacon Number,'
to be seven (four on average, and eight from all film actors and actresses of
any nationality)~\cite{kevinBacon,graphTheory2}.
Tjadan maintains a website called the {\em The Oracle of Bacon} at
http://www.cs.virginia.edu/oracle/ which provides
an interface to compute Bacon numbers.

Determining the impact the small-world phenomenon
has on the dynamic behavior of a distributed system is an open
research issue.  If small changes to
an edge set of a graph can have a dramatic impact on its global structural
properties, the same changes might affect its behavior as well.  This issue
was studied in the context of infectious diseases and games.  With
infectious diseases, shortest path length implies faster spreading
of the disease.  Therefore, global dynamics do depend on coupling
topology.  Games on graphs, such as the Prisoner's Dilemma reward game,
can be similarly analyzed.
When a low level of `hardness' in players exists, cooperation
dominates, but the timescale depends quite sensitively on the fraction
of shortcuts.  When the level of hardness in players is average, the
amount of cooperation which succeeds depends on the fraction of shortcuts.
Cooperation is more difficult to sustain when the number of shortcuts
increases due to those individuals who do not desire to
cooperate.  One would like to optimize both the spread and sustenance of
cooperation.

Another important open research question is whether individuals with only local
knowledge can effectively navigate in a small-world~\cite{small-worldNature}.
In addition to identifying \mbox{small-world} properties in the
acquaintance network of the
\mbox{United}~\mbox{States}, \mbox{Milgram's} experiment,
more importantly, illustrated that people with
only local knowledge of the network~(i.e.,~of their immediate acquaintances) were
actually successful at manually constructing acquaintance chains
of short length.
An open research question is if computers, as
opposed to humans, can construct short referral chains via an algorithmic
procedure.  
\mbox{Kleinberg} developed a decentralized algorithm which is capable of
constructing paths of short length in a \mbox{small-world} for one
particular model of small-world networks
which he developed~\cite{smallWorldPhen}.  Electronic
communications (e.g., e-mail) have made Milgram's experiment and
research results easily replicable.  For example, the Small World
project (http://smallworld.columbia.edu),
which is led by a group of researchers including Watts, is an online
experiment to test Milgram's six degrees of separation hypothesis.

Adamic tested the web and
discovered through experimentation that it is a small-world,
where vertices correspond to websites, as opposed to individual webpages,
and hyperlinks are considered to be undirected~\cite{smallWorld}.
She also addressed the implications these properties have on
searching the web and on discovering the structure of certain social communities
with a web presence.  $L$ was
estimated by averaging the paths in breath-first search (BFS)
trees.  $L$ was small (about 3.1
hops) and $C$ was 0.1078 compared to 2.3 $\times$ 10$^{\textrm{\tiny{-4}}}$
for a corresponding
random network with the same number of vertices and edges.
A second case study considered
directed links.  The largest SCC was computed on the graph.  BFS trees
were formed on a fraction of the vertices to sample the distribution
of distances.  $L$ was slightly higher due to the directed links and
$C$ was 0.081 compared to 1.05 $\times$ 10$^{\textrm{\tiny -3}}$
for a corresponding random graph
with the same number of vertices and edges.
This empirical evidence reveals that there is a
small-world network of websites.  A third
case study was performed on the .edu subgraph of the web which is
considerably smaller (i.e.,~distances between every vertex could be actually
computed).  Again, the largest SCC was computed.  $L$ was found to be 4.062
and $C$ was found to be 0.156 vs. 0.0012 for a corresponding
random graph with the same number of vertices and edges.
Hubs may foster short average path lengths between two randomly selected
(authority) webpages and thus may be integral to small-world properties in
the web.
`In summary, the largest SCCs of both sites in general and the subset of the
.edu sites are small-world networks with small average minimum
distances'~\cite{smallWorld}.

Adamic discusses how search engines could take advantage of these
small-world properties.
The idea of capitalizing on these structural features of the
web for web search is that it is more advantageous to return good
starting points, called `centers,' `index pages,' or `hubs,'
i.e.,~where `the distance from them to any other document within the group is on
average a minimum'~\cite{smallWorld}.
Search results thus can be grouped according to
such good starting points.  An application built around these ideas presents
a user with a list of centers sorted by the SCC size and allows the user
complete freedom in exploring these SCCs.

The next important research question is, aside from identification and
navigation: Can a small-world topology reveal clues to the structure of real
world communities, via the structure of the connections between documents
which members of those communities manually created?  Adamic
contends that a small-world topology can provide clues to the structure of
real-world communities.  She states that `exploring the link structure
of documents which belong to a particular topic could reveal the
underlying relationship between people and organizations'~\cite{smallWorld}.
Adamic discovered, via the application she developed,
that the pro-life community in the real world is
larger, more closely knit, and better organized than the pro-choice
community---by calculating the number of sites contained in each SCC.
Such an observation can have significant implications for marketing
strategies and red ribbon campaigns.

In summary, small-worlds present opportunities for recommender
systems.  If identified, not only do they help model users and communities
implicitly by revealing social structure, but also help connect people via
short chains. For example, if search engines could take advantage of the web's
small-world property,
then users with only local knowledge of the web may
actually be able to find and construct short paths between pairs of webpages.
Since \mbox{Albert}, \mbox{Hawoong}, and \mbox{Barab{\'{a}}si}
have shown the diameter of the web
to be nineteen~\cite{diameter}, if one knows
where one is going, then one can get there fast.
The {\em diameter} $d$ of a graph $G$ is the `shortest path between the most
distant vertices'~\cite{graphTheory2}.  Currently web search engines
do not exploit short path lengths between webpages.  Finding such
short paths within information abundant spaces, akin to using a compass,
reduces information overload and expedites recommendation.
In conclusion, the main point to take from this section is that knowledge of
the existence of
certain, typically social,
structures and properties~(e.g.,~connectivity, \mbox{bow-tie}, or
\mbox{small-worlds}) can be exploited by recommenders, such as
search engines, to intelligently provide more
effective results.

\section{Broadening Issues}
\label{broadening}

As we have illustrated, recommender systems are not
used in isolation but are rather cast in a broad social context.
In this section we briefly discuss broadening issues regarding
recommender systems, such as evaluation,
targeting, and privacy and trust.  This coverage of broadening
aspects of recommendation is not meant to be exhaustive.  While each
issue warrants survey in isolation, we only make some cursory remarks
here.  This section is intended to provide
pointers to authoritative sources to the reader interested in how
recommendation affects these topics.

\subsection{Evaluation}

While evaluation is critical to the success of a
recommender system or algorithm, rigorous evaluations are 
rarely performed, mainly due to the lack of universally
accepted formal methods for system evaluation.
Analyzing a recommendation algorithm from a functional perspective is
the most popular approach to evaluation~\cite{CFalgos}.
Such an approach typically involves measuring `predictive accuracy' via
a training/test set analysis~\cite{eigentaste}.
An alternate, more personal and social, approach is
to conduct HCI studies with user participants~\cite{RS-HCI}.
The former approach is employed more frequently than the latter.
These two approaches are at opposite ends of an evaluation spectrum
and are referred to as off-line vs. on-line evaluation~\cite{online-eval}.

\subsubsection*{Functional-oriented Evaluations}

Most evaluations of recommender algorithms
use standard IR metrics like precision and recall~\cite{PHOAKS}.
The GroupLens recommender systems research group has evaluated
recommendation algorithms for e-commerce~\cite{analysisRecAlgosEComm}.
They investigated traditional data mining techniques~(e.g.,
association rule mining in transactional data~\cite{miningAssociationRules}),
nearest-neighbor collaborative filtering, and dimensionality reduction.
Again, the evaluation metrics discussed are traditional:
support and confidence~\cite{miningAssociationRules},
and precision and recall~\cite{SaltonIRBook}.

After~\cite{recSys}, the second most highly cited article in the
recommender systems literature is a paper describing empirical
evaluation of the predictive accuracy of collaborative filtering
algorithms~\cite{CFalgos}.  Breese, Heckerman, and Kadie focus on
predictive accuracy rather than efficiency.  Their approach
to evaluation is purely statistical. They developed two classes of
evaluation metrics to characterize accuracy: average absolute deviation of
predictions and utility of ranked lists of recommended items.
Breese et al. compute these metrics across popular functional approaches to
collaborative filtering, such as correlation, Pearson's
$r$, vector similarity, inverse user frequency,
and statistical Bayesian methods---in
various ratings datasets, including the EachMovie dataset.
Although human satisfaction is not considered in this analysis,
the evaluation techniques introduced in~\cite{CFalgos} have become a de facto
standard for recommender systems~\cite{eigentaste}.

While there is nothing intrinsically bad about these functional-oriented
approaches to evaluation, they
do not capture the underlying
social process involved in recommendation.
As echoed repeatedly in this survey, recommendation is an inherently social
process and recommender systems ultimately connect people.
Evaluating a recommender system via a purely mathematical
analysis of its functional
approximation (predictive accuracy) ignores this integral
social process~\cite{online-eval}.  In addition, recommendations
must be explainable~\cite{explainableRecs}
and believable to users.   Functions are
not always explainable; a function could be a black
box to a user, which provides no transparency~\cite{explainableRecs}.
Therefore, traditional training/test set approaches~\cite{eigentaste}
and associated metrics,
such as precision and recall, and support and confidence, are
inadequate for algorithm evaluation from a social perspective.

\subsubsection*{Social-oriented Evaluations}

Swearingen and Sinha posit that the effectiveness of a recommender
system transcends the predictive accuracy of its underlying
algorithm~\cite{RS-HCI}.  The
opposite end of the recommendation evaluation spectrum is
purely social and entails conducting satisfaction surveys and studies.
Such studies are rare largely because of the high cost involved.
Sinha and \mbox{Swearingen} conducted
studies with users, as part of the HUBRIS (HUman Benchmarking of Recommender
Systems) project at UC Berkeley, directed toward
comparing recommendations given by friends to those by
six commercial and widely available recommender
systems~\cite{comparingevaluation}.  Their study also
addressed the degree to which assessments of overall recommender
system performance were correlated with the quality of the
recommendations or the user interface.  The ultimate objective of
their study is to develop a user-centered design
approach toward recommender systems.  Their initial results indicate that
an effective recommender inspires user trust in the system, provides
explainable recommendations w.r.t. system logic
(termed `transparency')~\cite{transparencyRS},
and makes serendipitous connections.

\subsubsection*{A Hybrid Approach}

In part to reconcile these two extreme approaches toward evaluation,
Mirza, Keller, and Ramakrishnan developed the Jumping Connections model
to evaluate recommendation
algorithms based on the number of people they connect.
For example, recommendation algorithm $A$ can be
compared to recommendation algorithm $B$ by the number of people
connected by the jump that $A$ models vs. the number
of people connected by the jump that $B$ models.
Such an approach places an emphasis on the underlying social
element to recommendation and the social network induced as
a result. 

The most striking contribution of their approach to evaluation
is that it resides at the fringe of HCI.  Using `the number of people
brought together' by a particular jump as an evaluation metric for
recommender systems is not as complex or sophisticated
as studying evaluation from a traditional functional-oriented
perspective, but is
more cost effective than an empirical study.
Moreover, it is 
not as social-oriented as a full blown study with user
participants.

\subsection{Targeting}

\begin{figure}
\centering
\begin{tabular}{c}
\includegraphics[width=8.25cm,height=5cm]{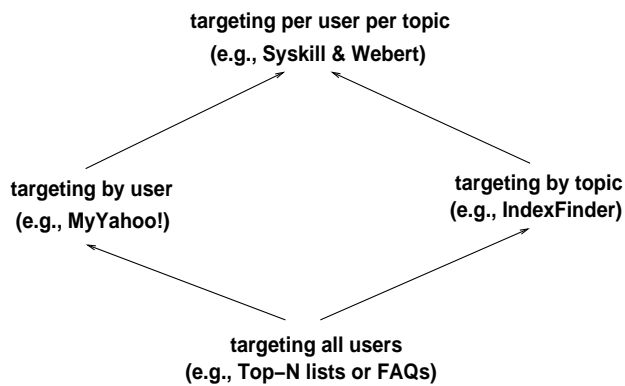}
\end{tabular}
\caption{Targeting a recommender system.
A lattice structure is induced by choices of targeting dimension:
to all users, a particular user, a topic, or per user per topic.}
\label{targeting}
\end{figure}

Targeting answers the question `for who or what are we building this
system?' and can be critical to the success of a recommender system.
Systems can be targeted to, e.g., all users, a particular user,
a set of topics, or per user per topic.
A lattice induced by these choices of targeting
dimensions is shown in Fig.~\ref{targeting}.
We briefly mention an example of each strategy in the targeting lattice.
Other possible targeting dimensions include geographic location or genre.

Top-$N$ lists and FAQs are designed to provide recommendations targeted to
all users~(Fig.~\ref{targeting}, bottom, the greatest lower bound).
`My Sites,' such as MyYahoo!~\cite{myYahooPers} are customizable web portals
targeted to the individual~(Fig.~\ref{targeting}---left).
The user selects areas of interest from several content
modules, such as news, stock quotes, weather, and sports, and the
system in turn provides recommended links to these topics. 
IndexFinder~\cite{adaptive} adapts websites based on various niche topics
(Fig.~\ref{targeting}---right).  The system automatically
synthesizes index webpages (i.e., hubs) consisting of links to other
webpages covering a particular common topic.
The synthesis of index pages is conducted by
mining user browsing patterns implicit in web logs and
conceptually clustering webpages to induce topics.
The `Syskill \& Webert' project~\cite{SyskillAndWebert} combines the
individual and topic targeting dimensions by targeting per user per
topic~(Fig.~\ref{targeting}, top, the least upper bound).
A user rates visited webpages on a three point scale and
the system induces a user profile by analyzing the
information on each webpage. The system then
recommends interesting websites to
users based on the learned profile and webpage topic. 

\subsection{Privacy and Trust}
\label{PrivacyAndTrust}

As expressed throughout this paper,
recommendation raises issues of privacy and trust~\cite{persPrivacy}.
Both an explicit and implicit approach to user modeling suffer from privacy
issues with the latter assuming more responsibility for evoking concern
due to its elusive nature.

A tradeoff exists between
collecting and leveraging as much user information as possible and
inspiring trust between the user and system.
Storing users' profiles and background leveraged for recommendation
on the client-side is an approach to empowering users with more control over their
personal information and degree of privacy. Users then control the
tradeoff between benefit and risk by deciding on their desired
level of involvement.  This sentiment is corroborated by
Singh, Yu, and Venkatraman~\cite{commBasedServiceLoc} who
compare community-based
networks and recommender systems to suggest that people really want to
control who sees their ratings and understand the process of
recommendation. Belkin however argues that with
sufficient reason to trust recommendations users are willing to give up
some measure of control and accept suggestions~\cite{dontKnow}.   Some
researchers even advocate exposing user profiles to build trust~\cite{trustRS}.
Other researchers contend that understanding why a recommendation was made
cultivates trust in a system~\cite{transparencyRS}.
Trust also can be designed into online experiences~\cite{designingTrust}.

The issue of user privacy
in general is not specific to a purely collaborative approach.  Privacy
and trust issues however are compounded in a collaborative setting as user
interests, background, or identity may be exposed to other users participating
in a social network in addition to the system.
Moreover, issues of privacy are not confined to a user and a recommender
system, but rather cascade across the social networks such systems
induce or identify.
Recall that when a recommender ratings dataset shatters, many isolated
communities take form.
As the hammock width
increases initially,
clusters tend to be fuzzy with many vertices bridging across
each cluster.
As the hammock width continues to increase, however, clusters
will crystallize with only one or two vertices serving as bridges across SCCs.
Often one or two users span such communities and
provide the basis for a serendipitous recommendation.  Such users or
bridge vertices are called `weak ties' in the social network
analysis community~\cite{SNA}.  Small-worlds are teeming with
weak ties which are critical to the small minimum average path length of the
network.
Weak ties help connect two or more strongly connected components.  
While weak ties are important to providing serendipitous
recommendations, just knowing that one exists (possibly by receiving an
unexpected recommendation) compromises the privacy of the person (the
weak tie) which
fostered the connection \cite{privacyRisks}.  For example, consider an
avid music fan who only rates CDs of Italian operas.  A recommendation of
Indian classical
music for such a user implies that there exists at least one individual spanning
these two communities.  The rare nature of the tie compromises identity.

\section{Future Work and Conclusion}
\label{future}

The field of recommender systems is young and evolving.
We have identified some directions for future work.

\begin{itemize}

\item {\bf Distributed Recommendation Infrastructure and
Interoperability}: Taking recommendation out of specific systems and
casting it in a broader, distributed information
infrastructure is a direction for future research~\cite{CliffordLynch}.
Such an infrastructure~\cite{broaderPers} fosters the possibility of users
managing and maintaining their own client-side
user model and context to share at their discretion with participating
recommenders.
Addressing and resolving technical issues, such as interoperability and
standardization, and social issues, such as buy-in, is essential to the
realization of this vision.

\item {\bf Formal Recommendation Modeling and Design Methodologies}:
As with any young, evolving, and multidisciplinary field,
recommendation is lacking
unified methodologies to study, design, and evaluate systems.
Without such methodologies, unsystematic
development will continue to persist
resulting in kludge with consequential problems of cost and
interoperability.  Initial
steps toward such models are reflected in projects such as
Jumping Connections~\cite{jumpingConnections} and the on-line
evaluation framework~\cite{online-eval}.
Design patterns and languages may help capture solutions to recurrent modeling
problems, such as sparsity and privacy,
and foster the semi-automatic construction of systems.

\item {\bf An HCI Approach---Designing for Interaction}:
Developing and integrating less intrusive and salient interfaces for explicit
product evaluations and ratings with recommendation delivery
interfaces is a compelling line of future research.
The interest and need for such research, including possibly usability
evaluations, on extant and new streamlined UIs for recommenders systems, is
reflected in a recent {\em ACM Transactions of Computer-Human Interaction}
call for papers~\cite{TOCHI-RS}.
Furthermore, studying user interactions with recommender systems is becoming
an increasingly popular way to design such systems~\cite{interactionRS}.
HCI researchers involved in this effort are optimistic that results
of such studies will lead to general design guidelines.
These two directions present opportunities for the HCI community
to make inroads to recommender systems research which needs such expertise.

\item {\bf Information Appliances}: Computing is becoming more and more
ubiquitous.  Physical computing devices and `information appliances,'
computer-enhanced
devices dedicated to a restricted suite of information-oriented tasks, are
no longer a vision, but rather a reality~\cite{InformationAppliances}.
The scope of information appliances is no longer limited to
handheld computers, PDAs, and mobile phones, but has extended to
MP3 players and watches.
In addition, ubiquity is enriched by
voice applications and portals (e.g., Tellme, http://\hskip0ex www.\hskip0ex 
tellme.\hskip0ex com),
collectively called the `voice web'~\cite{speechRec},
avatars~\cite{webCompanions}, and
information kiosks~\cite{HermitageMuseum}.  These devices
present compelling opportunities to deliver `recommendations on demand.'
For example, consider delivering a restaurant
recommendation and associated coupon on a cell phone to
a businessman in search of a vegetarian meal while waiting in an airport. 
Leveraging the ubiquity of such notification systems for user modeling
and recommendation delivery is an open research issue.
 
\end{itemize}

In conclusion, we have provided a survey of recommender systems research
according to the way they model users and resulting connections they
achieve or identify.  We feel this is a more holistic
approach to recommendation as it captures the underlying social
element in all recommenders.
The reader will have noticed that intelligent techniques
to model users are slowly being augmented with approaches to exploit the
combinatorial social structure implicit in usage data.
The future of recommender systems will ultimately lie in mediating these
two approaches and developing unified methodologies to
systematize the process of representing users and building systems.

\acknowledgements
We acknowledge the suggestions of the three anonymous referees whose
comments have improved this article's presentation.
We thank \mbox{Naren}~\mbox{Ramakrishnan} for planting the idea for
this survey
during his Spring~2001 offering of CS6604: Recommender Systems at
Virginia Tech.  In addition,
we thank him for feedback regarding initial drafts of this
survey.  Furthermore, discussion summaries and feedback from those in
CS6604 improved the organization of this paper.
We also thank Padmapriya Kandhadai, Batul Mirza, Cal Ribbens, and
Chad Wingrave at Virginia Tech for providing constructive comments.

\bibliographystyle{my-alpha}
\bibliography{RSsurvey}

\newcommand{\etalchar}[1]{$^{#1}$}
\begin{thebibliography}{AWWY99}

\bibitem[Ada99]{smallWorld}
L.~A. Adamic.
\newblock {The Small World Web}.
\newblock In S.~Abiteboul and A-M. Vercoustre, editors, {\em Proceedings of the
  European Conference on Digital Libraries (ECDL'99)}, Lecture Notes in
  Computer Science, pages 443--452, Paris, France, September 1999. Springer.

\bibitem[AIS93]{miningAssociationRules}
R.~Agrawal, T.~Imielinski, and A.~N. Swami.
\newblock {Mining Association Rules between Sets of Items in Large Databases}.
\newblock In P.~Buneman and S.~Jajodia, editors, {\em Proceedings of the {ACM}
  International Conference on Management of Data (SIGMOD'93)}, pages 207--216,
  Washington, DC, May 1993. ACM Press.

\bibitem[AJB99]{diameter}
R.~Albert, H.~Jeong, and A.-L. Barab{\'{a}}si.
\newblock {The Diameter of the World Web Web}.
\newblock {\em Nature}, Vol. 401:pages 130--131, September 1999.

\bibitem[AKK98]{movie-content}
J.~Alspector, A.~Kolez, and N.~Karunanithi.
\newblock {Comparing Feature-Based and Clique-Based User Models for Movie
  Selection}.
\newblock In {\em Proceedings of the Third ACM Conference on Digital
  Libraries}, pages 11--18, Pittsburgh, PA, June 1998. ACM Press.

\bibitem[AR02]{webCompanions}
E.~Andr{\'{e}} and T.~Rist.
\newblock {From Adaptive Hypertext to Personalized Web Companions}.
\newblock {\em Communications of the ACM}, Vol. 45(5):pages 43--46, May 2002.

\bibitem[ASBS00]{classesSmallWorld}
L.~A.~N. Amaral, A.~Scala, M.~Barth{\'{e}}l{\'{e}}my, and H.~E. Stanley.
\newblock {Classes of Behavior of Small-World Networks}.
\newblock In {\em Proceedings of the National Academy of Science, USA},
  Vol.~97, pages 11149--11152, January 2000.

\bibitem[AT99]{miningPers}
G.~Adomavicius and A.~Tuzhilin.
\newblock {User Profiling in Personalization Applications through Rule
  Discovery and Validation}.
\newblock In {\em Proceedings of the Fifth ACM SIGKDD International Conference
  on Knowledge Discovery and Data Mining (KDD'99)}, pages 377--381, San Diego,
  CA, August 1999. ACM Press.

\bibitem[AT01]{mutliDRS}
G.~Adomavicius and A.~Tuzhilin.
\newblock {Multidimensional Recommender Systems: A Data Warehousing Approach}.
\newblock In L.~Fiege, G.~M{\"{u}}hl, and U.~G. Wilhelm, editors, {\em Second
  International Workshop on Electronic Commerce (WELCOM'01)}, Vol. 2232 of {\em
  Lecture Notes in Computer Science}, pages 180--192, Heidelberg, Germany,
  November 2001. Springer-Verlag.

\bibitem[ATH00]{authorityQuality}
B.~Amento, L.~Terveen, and W.~Hill.
\newblock {Does ``Authority'' Mean Quality? Predicting Expert Quality Ratings
  of Web Documents}.
\newblock In {\em Proceedings of the Twenty-third Annual International ACM
  Conference on Research and Development in Information Retrieval (SIGIR'00)},
  pages 296--303, Athens, Greece, July 2000. ACM Press.

\bibitem[AWWY99]{horting}
C.~C. Aggarwal, J.~L. W., K.~Wu, and P.~S. Yu.
\newblock {Horting Hatches an Egg: A Graph-Theoretic Approach to Collaborative
  Filtering}.
\newblock In {\em Proceedings of the Fifth ACM SIGKDD International Conference
  on Knowledge Discovery and Data Mining (KDD'99)}, pages 201--212, San Diego,
  CA, August 1999. ACM Press.

\bibitem[AZ97]{provisionRec}
C.~Avery and R.~Zeckhauser.
\newblock {Recommender Systems for Evaluating Computer Messages}.
\newblock {\em Communications of the ACM}, Vol. 40(3):pages 88--89, March 1997.

\bibitem[BA99]{Barabasi}
A.-L. Barab{\'{a}}si and R.~Albert.
\newblock {Emergence of Scaling in Random Networks}.
\newblock {\em Science}, Vol. 286:pages 509--512, October 1999.

\bibitem[Bau99]{joinCBFandCF}
P.~Baudisch.
\newblock {Joining Collaborative and Content-based Filtering}.
\newblock In {\em Proceedings of the ACM CHI Workshop on Interacting with
  Recommender Systems}, Pittsburgh, PA, May 1999. ACM Press.

\bibitem[BC92]{irif}
N.~J. Belkin and W.~B. Croft.
\newblock {Information Filtering and Information Retrieval: Two Sides of the
  Same Coin?}
\newblock {\em Communications of the ACM}, Vol. 35(12):pages 29--38, December
  1992.

\bibitem[BDO95]{LSI}
M.~W. Berry, S.~T. Dumais, and G.~W. O'Brien.
\newblock {Using Linear Algebra for Intelligent Information Retrieval}.
\newblock {\em SIAM Review}, Vol. 37(4):pages 573--595, January--February 1995.

\bibitem[Bel00]{dontKnow}
N.~J. Belkin.
\newblock {Helping People Find What They Don't Know}.
\newblock {\em Communications of the ACM}, Vol. 43(8):pages 58--61, August
  2000.

\bibitem[Ber00]{InformationAppliances}
E.~Bergman, editor.
\newblock {\em {Information Appliances and Beyond}}.
\newblock The Morgan Kaufmann Series on Interactive Technologies. Morgan
  Kaufmann, San Francisco, CA, 2000.

\bibitem[Ber01]{causticCookies}
H.~Berghel.
\newblock {Caustic Cookies}.
\newblock {\em Communications of the ACM}, Vol. 44(5):pages 19--22, May 2001.

\bibitem[BH01]{multipleSources}
C.~Basu and H.~Hirsh.
\newblock {Using Multiple Information Sources for Recommendation}.
\newblock In {\em Proceedings of the Twenty-fourth Annual International ACM
  SIGIR Conference, Workshop on Recommender Systems}, New Orleans, LA, November
  2001. ACM Press.

\bibitem[BHK98]{CFalgos}
J.~S. Breese, D.~Heckerman, and C.~Kadie.
\newblock {Empirical Analysis of Predictive Algorithms for Collaborative
  Filtering}.
\newblock In {\em Proceedings of the Fourteenth Annual Conference on
  Uncertainty in Artificial Intelligence}, pages 43--52, Madison, WI, July
  1998.

\bibitem[BKM{\etalchar{+}}00]{bow-tie}
A.~Broder, R.~Kumar, F.~Maghoul, P.~Raghavan, S.~Rajagopalan, R.~Stata,
  A.~Tomkins, and J.~Wiener.
\newblock {Graph Structure in the Web}.
\newblock In {\em Proceedings of the Ninth International World Wide Web
  Conference (WWW9)}, Amsterdam, Netherlands, May 2000.

\bibitem[BMM96]{machineLearningUserProfiles}
E.~Bloedorn, I.~Mani, and T.~R. MacMillan.
\newblock {Representational Issues in Machine Learning of User Profiles}.
\newblock In {\em AAAI Spring Symposium on Machine Learning in Information
  Access (MLIA)}, Stanford, CA, March 1996. AAAI Press.

\bibitem[BP98]{anatomy}
S.~Brin and L.~Page.
\newblock {The Anatomy of a Large-Scale Hypertextual {Web} Search Engine}.
\newblock In {\em Proceedings of the Seventh International World Wide Web
  Conference (WWW7)}, Brisbane, Australia, April 1998. Elsevier Science.

\bibitem[Bro03]{SIGIR2003keynote}
A.~Broder.
\newblock {Exploring, Modeling, and Using the Web Graph}.
\newblock Keynote to the Twenty-six Annual International ACM Conference on
  Research and Development in Information Retrieval (SIGIR'03), July 2003.

\bibitem[BS97]{Fab}
M.~Balabanovi{\'{c}} and Y.~Shoham.
\newblock {Fab: Content-Based, Collaborative Recommendation}.
\newblock {\em Communications of the ACM}, Vol. 40(3):pages 66--72, March 1997.

\bibitem[Bur99]{banana}
R.~Burke.
\newblock {Integrating Knowledge-Based and Collaborative Filtering Recommender
  Systems}.
\newblock In {\em Proceedings of the Workshop on Artificial Intelligence for
  Electronic Commerce}, pages 69--72, Orlando, FL, July 1999. AAAI Press.

\bibitem[Bus45]{AsWeMayThink}
V.~Bush.
\newblock {As We May Think}.
\newblock {\em The Atlantic Monthly}, Vol. 176(1):pages 101--108, July 1945.

\bibitem[CBLW01]{inferring-user-interest}
M.~Claypool, D.~Brown, P.~Le, and M.~Waseda.
\newblock {Inferring User Interest}.
\newblock {\em IEEE Internet Computing}, Vol. 5(6):pages 32--39,
  November--December 2001.

\bibitem[CD00]{power-law}
J.~M. Carlson and J.~Doyle.
\newblock {Highly Optimized Tolerance: A Mechanism for Power Laws in Designed
  Systems}.
\newblock Technical report, California Institute of Technology, 2000.

\bibitem[CDA00]{broaderPers}
I.~Cingil, A.~Dogac, and A.~Azgin.
\newblock {A Broader Approach to Personalization}.
\newblock {\em Communications of the ACM}, Vol. 43(8):pages 136--141, August
  2000.

\bibitem[CDK{\etalchar{+}}99]{miningLinkStr}
S.~Chakrabarti, B.~E. Dom, S.~R. Kumar, P.~Raghavan, S.~Rajagopalan,
  A.~Tomkins, D.~Gibson, and J.~Kleinberg.
\newblock {Mining the Web's Link Structure}.
\newblock {\em IEEE Computer}, Vol. 32(8):pages 60--67, August 1999.

\bibitem[CGM{\etalchar{+}}99]{early-rater}
M.~Claypool, A.~Gokhale, T.~Miranda, P.~Murnikov, D.~Netes, and M.~Sartin.
\newblock {Combining Content-Based and Collaborative Filters in an Online
  Newspaper}.
\newblock In {\em Proceedings of the Twenty-second Annual International ACM
  SIGIR Conference, Workshop on Recommender Systems}, Berkeley, CA, August
  1999. ACM Press.

\bibitem[CR96]{BEV}
J.~M. Carroll and M.~B. Rosson.
\newblock {Developing the Blacksburg Electronic Village}.
\newblock {\em Communications of the ACM}, Vol. 39(12):pages 69--74, December
  1996.

\bibitem[Den82]{electronicJunk}
P.~Denning.
\newblock {Electronic Junk}.
\newblock {\em Communications of the ACM}, Vol. 25(3):pages 163--165, March
  1982.

\bibitem[ER60]{Erdos}
P.~Erd{\H{o}}s and A.~R{\'{e}}nyi.
\newblock {On the Evolution of Random Graphs}.
\newblock {\em Publications of the Mathematical Institute of the Hungarian
  Academy of Sciences}, Vol. 5:pages 17--61, 1960.

\bibitem[FD92]{lsiCACM}
P.~W. Foltz and S.~T. Dumais.
\newblock {Personalized Information Delivery: An Analysis of Information
  Filtering Methods}.
\newblock {\em Communications of the ACM}, Vol. 35(12):pages 51--60, December
  1992.

\bibitem[FISS98]{effusivity}
Y.~Fruend, R.~Iyer, R.~Schapire, and Y.~Singer.
\newblock {An Efficient Boosting Algorithm for Combining Preferences}.
\newblock In {\em Proceedings of the Fifteenth International Conference on
  Machine Learning}, pages 170--178, Madison, WI, July 1998. Morgan Kaufmann.

\bibitem[FLGC02]{selfOrg}
G.~W. Flake, S.~Lawrence, C.~L. Giles, and F.~M. Coetzee.
\newblock {Self-Organization and Identification of Web Communities}.
\newblock {\em IEEE Computer}, Vol. 35(3):pages 66--67, March 2002.

\bibitem[GKR99]{RSinterfaces}
A.~Grasso, M.~Koch, and A.~Rancati.
\newblock {Augmenting Recommender Systems by Embedding Interfaces into
  Practices}.
\newblock In {\em Proceedings of the International ACM SIGGROUP Conference on
  Supporting Group Work (GROUP'99)}, pages 267--275, Phoenix, AZ, November
  1999. ACM Press.

\bibitem[GNOT92]{cf}
D.~Goldberg, D.~Nichols, B.~M. Oki, and D.~Terry.
\newblock {Using Collaborative Filtering to Weave an Information Tapestry}.
\newblock {\em Communications of the ACM}, Vol. 35(12):pages 61--70, December
  1992.

\bibitem[GRGP00]{eigentaste}
K.~Goldberg, T.~Roeder, D.~Gupta, and C.~Perkins.
\newblock {Eigentaste: A Constant Time Collaborative Filtering Algorithm}.
\newblock Technical Report M00/41, Electronic Research Laboratory, University
  of California, Berkeley, August 2000.

\bibitem[GS00]{Goecks00}
J.~Goecks and J.~Shavlik.
\newblock {Learning Users' Interests by Unobtrusively Observing their Normal
  Behavior}.
\newblock In {\em Proceedings of the 2000 International Conference on
  Intelligent User Interfaces (IUI'00)}, Observing User Behavior, pages
  129--132, New Orleans, LA, January 2000. ACM Press.

\bibitem[Hay00a]{graphTheory1}
B.~Hayes.
\newblock {Graph Theory in Practice: Part I}.
\newblock {\em American Scientist}, Vol. 88(1):pages 9--13, January--February
  2000.

\bibitem[Hay00b]{graphTheory2}
B.~Hayes.
\newblock {Graph Theory in Practice: Part II}.
\newblock {\em American Scientist}, Vol. 88(2):pages 104--109, March--April
  2000.

\bibitem[HK96]{keyword}
E.~Housman and E.~Kaskela.
\newblock {State of the Art in Selective Dissemination of Information}.
\newblock In {\em Proceedings of the IEEE Transaction on Engineering and
  Writing Speech}, pages 100--112, Stanford, CA, March 1996. AAAI Press.

\bibitem[HKR00]{explainableRecs}
J.~Herlocker, J.~A. Konstan, and J.~Riedl.
\newblock {Explaining Collaborative Filtering Recommendations}.
\newblock In {\em Proceedings of the ACM Conference on Computer Supported
  Cooperative Work (CSCW'00)}, pages 241--250, Philadelphia, PA, December 2000.
  ACM Press.

\bibitem[HMAC02]{online-eval}
C.~Hayes, P.~Massa, P.~Avesani, and P.~Cunningham.
\newblock {An On-line Evaluation Framework for Recommender Systems}.
\newblock Technical Report TCD-CS-2002-19, Department of Computer Science,
  Trinity College Dublin, April 2002.

\bibitem[HT96]{communitySortedSearch}
W.~Hill and L.~Terveen.
\newblock {Using Frequency-of-Mention in Public Conversations for Social
  Filtering}.
\newblock In {\em Proceedings of the ACM Conference on Computer Supported
  Cooperative Work}, pages 106--112, Boston, MA, November 1996. ACM Press.

\bibitem[KB96]{content-based}
B.~Krulwich and C.~Burkley.
\newblock {Learning User Information Interests Through Extraction of
  Semantically Significant Phrases}.
\newblock In {\em Proceedings of the AAAI Spring Symposium on Machine Learning
  in Information Access}, pages 100--112, Stanford, CA, March 1996. AAAI Press.

\bibitem[KKR{\etalchar{+}}99]{WebGraph}
J.~Kleinberg, S.~R. Kumar, P.~Raghavan, S.~Rajagopalan, and A.~Tomkins.
\newblock {The Web as a Graph: Measurements, Models and Methods}.
\newblock In {\em Proceedings of the International Conference on Combinatorics
  and Computing}, 1999.

\bibitem[KL01]{WebStrucScience}
J.~Kleinberg and S.~Lawrence.
\newblock {The Structure of the Web}.
\newblock {\em Science}, Vol. 294:pages 1849--1850, November 2001.

\bibitem[Kle99]{KleinbergAuthoritative}
J.~Kleinberg.
\newblock {Authoritative Sources in a Hyperlinked Environment}.
\newblock {\em Journal of the ACM}, Vol. 46(5):pages 604--632, September 1999.

\bibitem[Kle00a]{small-worldNature}
J.~Kleinberg.
\newblock {Navigation in a Small World}.
\newblock {\em Nature}, Vol. 406:page 845, August 2000.

\bibitem[Kle00b]{smallWorldPhen}
J.~Kleinberg.
\newblock {The Small-World Phenomenon: An Algorithmic Perspective}.
\newblock In {\em Proceedings of the Thirty-second ACM Symposium on Theory of
  Computing (STOC'00)}, pages 163--170, Portland, OR, 2000. ACM Press.

\bibitem[KMM{\etalchar{+}}97]{GroupLens}
J.~A. Konstan, B.~N. Miller, D.~Maltz, J.~L. Herlocker, L.~R. Gordon, and
  J.~Riedl.
\newblock {GroupLens: Applying Collaborative Filtering to Usenet News}.
\newblock {\em Communications of the ACM}, Vol. 40(3):pages 77--87, March 1997.

\bibitem[KRRT99]{trawling}
R.~Kumar, P.~Raghavan, S.~Rajagopalan, and A.~Tomkins.
\newblock {Trawling the Web for Emerging Cyber-Communities}.
\newblock In {\em Proceedings of the Eighth International World Wide Web
  Conference (WWW8)}, Toronto, Canada, May 1999.

\bibitem[KSS97a]{ReferralWeb}
H.~Kautz, B.~Selman, and M.~Shah.
\newblock {Referral Web: Combining Social Networks and Collaborative
  Filtering}.
\newblock {\em Communications of the ACM}, Vol. 40(3):pages 63--65, March 1997.

\bibitem[KSS97b]{HiddenWeb}
H.~Kautz, B.~Selman, and M.~Shah.
\newblock {The Hidden Web}.
\newblock {\em AI Magazine}, Vol. 18(2):pages 27--36, 1997.

\bibitem[LSY03]{Amazon}
G.~Linden, B.~Smith, and J.~York.
\newblock {Amazon.com Recommendations: Item to Item Collaborative Filtering}.
\newblock {\em IEEE Internet Computing}, Vol. 7(1):pages 76--80,
  January--February 2003.

\bibitem[LT92]{if}
S.~Loeb and D.~Terry.
\newblock {Information Filtering}.
\newblock {\em Communications of the ACM}, Vol. 35(12):pages 26--28, December
  1992.

\bibitem[Lyn01]{CliffordLynch}
C.~Lynch.
\newblock {Personalization and Recommender Systems in the Larger Context: New
  Directions and Research Questions (Keynote Speech)}.
\newblock In {\em Proceedings of the Joint DELOS-NSF Workshop on
  Personalisation and Recommender Systems in Digital Libraries}, pages 84--88,
  Dublin, Ireland, 2001.

\bibitem[MAB00]{WebMining}
M.~D. Mulvenna, S.~S. Anand, and A.~G. B{\"{u}}chner.
\newblock {Personalization on the Net using Web Mining}.
\newblock {\em Communications of the ACM}, Vol. 43(8):pages 122--125, August
  2000.

\bibitem[MBG{\etalchar{+}}01]{HermitageMuseum}
F.~Mintzer, G.~W. Braudaway, F.~P. Giordano, J.~C. Lee, Karen~A. Magerlein,
  S.~D'Auria, A.~Ribak, G.~Shapir, F.~Schiattarella, J.~Tolva, and A.~Zelenkov.
\newblock {Populating the Hermitage Museum's New Web Site}.
\newblock {\em Communications of the ACM}, Vol. 44(8):pages 52--60, August
  2001.

\bibitem[MCS00]{autoPersWebMining}
B.~Mobashier, R.~Cooley, and J.~Srivastava.
\newblock {Automatic Personalization Based on Web Usage Mining}.
\newblock {\em Communications of the ACM}, Vol. 43(8):pages 142--151, August
  2000.

\bibitem[ME95]{cold-start}
D.~Maltz and K.~Ehrlich.
\newblock {Pointing the Way: Active Collaborative Filtering}.
\newblock In {\em Proceedings of the ACM Conference on Human Factors in
  Computing Systems (CHI'95)}, pages 202--209, Denver, CO, May 1995. ACM Press.

\bibitem[Mil67]{Milgram}
S.~Milgram.
\newblock {The Small World Problem}.
\newblock {\em Psychology Today}, Vol. 1(61):pages 56--58, 1967.

\bibitem[Mir01]{BatulThesis}
B.~J. Mirza.
\newblock {Jumping Connections: A Graph-Theoretic Model for Recommender
  Systems}.
\newblock Master's thesis, Virginia Tech, February 2001.
\newblock Available at
  http://scholar.lib.vt.edu/theses/available/etd-02282001-175040/.

\bibitem[MKR03]{jumpingConnections}
B.~J. Mirza, B.~J. Keller, and N.~Ramakrishnan.
\newblock {Studying Recommendation Algorithms by Graph Analysis}.
\newblock {\em Journal of Intelligent Information Systems}, Vol. 20(2):pages
  131--160, 2003.

\bibitem[MMLP97]{SIFTER}
J.~Mostafa, S.~Mukhopadhyay, W.~Lam, and M.~Palakal.
\newblock {A Multilevel Approach to Intelligent Information Filtering: Model,
  System, and Evaluation}.
\newblock {\em ACM Transactions on Information Systems}, Vol. 15(4):pages
  368--399, October 1997.

\bibitem[MPR00]{myYahooPers}
U.~Manber, A.~Patel, and J.~Robinson.
\newblock {Experience with Personalization on Yahoo!}
\newblock {\em Communications of the ACM}, Vol. 43(8):pages 35--39, August
  2000.

\bibitem[MR00]{bookContent}
R.~Mooney and L.~Roy.
\newblock {Content-Based Book Recommending Using Learning for Text
  Categorization}.
\newblock In {\em Proceedings of the Fifth ACM Conference on Digital
  Libraries}, pages 195--204, San Antonio, TX, July 2000. ACM Press.

\bibitem[NM01]{bioDL}
C.~Nevill-Manning.
\newblock {The Biological Digital Library}.
\newblock {\em Communications of the ACM}, Vol. 44(5):pages 41--42, May 2001.

\bibitem[Osa96]{bibliometrics}
F.~Osareh.
\newblock {Bibliometrics, Citation Analysis and Co-Citation Analysis: A Review
  of Literature I}.
\newblock {\em Libri}, Vol. 46:pages 149--158, 1996.

\bibitem[Pap01]{Papatheodorou}
C.~Papatheodorou.
\newblock {Machine Learning in User Modeling}.
\newblock {\em Lecture Notes in Computer Science}, Vol. 2049:pages 286--294,
  September 2001.

\bibitem[PB97]{Pazzani}
M.~J. Pazzani and D.~Billsus.
\newblock {Learning and Revising User Profiles: The Identification of
  Interesting Web Sites}.
\newblock {\em Machine Learning}, Vol. 27(3):pages 313--331, 1997.

\bibitem[PE00]{adaptive}
M.~Perkowitz and O.~Etzioni.
\newblock {Adaptive Web Sites}.
\newblock {\em Communications of the ACM}, Vol. 43(8):pages 152--158, August
  2000.

\bibitem[PMB96]{SyskillAndWebert}
M.~Pazzani, J.~Muramatsu, and D.~Billsus.
\newblock {Syskill and Webert: Identifying Interesting Web Sites}.
\newblock In {\em Proceedings of the Thirteenth National Conference on
  Artificial Intelligence (AAAI-96)}, pages 54--61, Portland, OR, August 1996.
  AAAI Press.

\bibitem[PR03]{piis}
S.~Perugini and N.~Ramakrishnan.
\newblock {Personalizing Interactions with Information Systems}.
\newblock {\em Advances in Computers}, Vol. 57: Information Repositories:pages
  323--382, September 2003.
\newblock Invited contribution.

\bibitem[Rie00]{persViewsOfPers}
D.~Riecken.
\newblock {Personalized Views of Personalization}.
\newblock {\em Communications of the ACM}, Vol. 43(8):pages 27--28, August
  2000.

\bibitem[Rie01]{persPrivacy}
J.~Riedl.
\newblock {Personalization and Privacy}.
\newblock {\em IEEE Internet Computing}, Vol. 5(6):pages 29--31,
  November-December 2001.

\bibitem[RIS{\etalchar{+}}94]{GroupLensOrg}
P.~Resnick, N.~Iacovou, M.~Sushak, P.~Bergstrom, and J.~Riedl.
\newblock {GroupLens: An Open Architecture for Collaborative Filtering of
  Netnews}.
\newblock In {\em Proceedings of the ACM Conference on Computer Supported
  Cooperative Work (CSCW'94)}, pages 175--186, Chapel Hill, NC, October 1994.
  ACM Press.

\bibitem[RKM{\etalchar{+}}01]{privacyRisks}
N.~Ramakrishnan, B.~J. Keller, B.~J. Mirza, A.~Y. Grama, and G.~Karypis.
\newblock {Privacy Risks in Recommender Systems}.
\newblock {\em IEEE Internet Computing}, Vol. 5(6):pages 54--62,
  November--December 2001.

\bibitem[RN95]{AIMA}
S.~Russell and P.~Norvig.
\newblock {\em {Artificial Intelligence: A Modern Approach}}.
\newblock Prentice Hall Series in Artificial Intelligence. Prentice Hall, Upper
  Saddle River, NJ, 1995.

\bibitem[Roc71]{relevanceFeedback}
J.~J. Rocchio.
\newblock {Relevance Feedback in Information Retrieval}.
\newblock In G.~Salton, editor, {\em {The SMART Retrieval System: Experiments
  in Automatic Document Processing}}, pages 313--323. Prentice-Hall, Englewood
  Cliffs, NJ, 1971.

\bibitem[RP97]{Siteseer}
J.~Rcuker and M.~J. Polano.
\newblock {Siteseer: Personalized Navigation for the Web}.
\newblock {\em Communications of the ACM}, Vol. 40(3):pages 73--75, March 1997.

\bibitem[RV97]{recSys}
P.~Resnick and H.~R. Varian.
\newblock {Recommender Systems}.
\newblock {\em Communications of the ACM}, Vol. 40(3):pages 56--58, March 1997.

\bibitem[RZFK00]{repSys}
P.~Resnick, R.~Zeckhauser, E.~Friedman, and K.~Kuwabara.
\newblock {Reputation Systems}.
\newblock {\em Communications of the ACM}, Vol. 43(12):pages 45--48, December
  2000.

\bibitem[SB98]{RI}
R.~S. Sutton and A.~G. Barto.
\newblock {\em {Reinforcement Learning: An Introduction}}.
\newblock Adaptive Computation and Machine Learning. MIT Press, Cambridge, MA,
  1998.

\bibitem[SB02]{speechRec}
S.~Srinivasan and E.~Brown.
\newblock {Is Speech Recognition Becoming Mainstream?}
\newblock {\em IEEE Computer}, Vol. 35(4):pages 38--41, April 2002.

\bibitem[SCDT00]{WebUsageMining}
J.~Srivastava, R.~Cooley, M.~Deshpande, and P.-N. Tan.
\newblock {Web Usage Mining: Discovery and Applications of Usage Patterns from
  Web Data}.
\newblock {\em SIGKDD Explorations}, Vol. 1(2):pages 12--23, January 2000.

\bibitem[Sha94]{Shardanand}
U.~Shardanand.
\newblock {\em {Social Information Filtering for Music Recommendation}}.
\newblock {Ph.D.} dissertation, Massachusetts Institute of Technology, 1994.

\bibitem[Shn00]{designingTrust}
B.~Shneiderman.
\newblock {Designing Trust into Online Experiences}.
\newblock {\em Communications of the ACM}, Vol. 43(12):pages 57--59, December
  2000.

\bibitem[SKB{\etalchar{+}}98]{GroupLens2}
B.~Sarwar, J.~Konstan, J.~Borchers, A.~Herlocker, J.~Miller, and J.~Riedl.
\newblock {Using Filtering Agents to Improve Prediction Quality in the
  GroupLens Research Collaborative Filtering System}.
\newblock In {\em Proceedings of the ACM Conference on Computer Supported
  Cooperative Work (CSCW'98)}, pages 345--354, Seattle, WA, November 1998. ACM
  Press.

\bibitem[SKKR00]{analysisRecAlgosEComm}
B.~Sarwar, G.~Karypis, J.~Konstan, and J.~Riedl.
\newblock {Analysis of Recommendation Algorithms for E-Commerce}.
\newblock In {\em Proceedings of the Second ACM Conference on Electronic
  Commerce}, pages 158--167, Minneapolis, MN, 2000. ACM Press.

\bibitem[SKR99]{recEComm}
J.~B. Schafer, J.~A. Konstan, and J.~Riedl.
\newblock {Recommender Systems in E-Commerce}.
\newblock In {\em Proceedings of the First ACM Conference on Electronic
  Commerce}, pages 158--166, Denver, CO, November 1999. ACM Press.

\bibitem[SM83]{SaltonIRBook}
G.~Salton and M.~J. McGill.
\newblock {\em {Introduction to Modern Information Retrieval}}.
\newblock McGraw-Hill Computer Science Series. McGraw-Hill, New York, 1983.

\bibitem[SM95]{socialFiltering}
U.~Shardanand and P.~Maes.
\newblock {Social Information Filtering: Algorithms for Automating ``Word of
  Mouth''}.
\newblock In {\em Proceedings of the ACM Conference on Human Factors in
  Computing Systems (CHI'95)}, pages 210--217, Denver, CO, May 1995. ACM Press.

\bibitem[SN99]{hybrid}
I.~Soboroff and C.~Nicholas.
\newblock {Combining Content and Collaboration in Text Filtering}.
\newblock In {\em Proceedings of the IJCAI'99 Workshop on Machining Learning in
  Information Filtering}, pages 86--91, Stockholm, Sweden, August 1999.

\bibitem[Spi00]{evaluation}
M.~Spiliopoulou.
\newblock {Web Usage Mining for Web Site Evaluation}.
\newblock {\em Communications of the ACM}, Vol. 43(8):pages 127--134, August
  2000.

\bibitem[SS01a]{comparingevaluation}
R.~Sinha and K.~Swearingen.
\newblock {Comparing Recommendations Made by Online Systems and Friends}.
\newblock In {\em Proceedings of the Joint DELOS-NSF Workshop on
  Personalisation and Recommender Systems in Digital Libraries}, pages 73--78,
  Dublin, Ireland, June 2001.

\bibitem[SS01b]{RS-HCI}
K.~Swearingen and R.~Sinha.
\newblock {Beyond Algorithms: An HCI perspective on Recommender Systems}.
\newblock In {\em Proceedings of the Twenty-fourth Annual International ACM
  SIGIR Conference, Workshop on Recommender Systems}, New Orleans, LA, November
  2001. ACM Press.

\bibitem[SS02a]{transparencyRS}
R.~Sinha and K.~Swearingen.
\newblock {The Role of Transparency in Recommender Systems}.
\newblock In {\em Proceedings of the ACM Conference on Human Factors in
  Computing Systems (CHI'02)}, pages 830--831, Minneapolis, MN, April 2002. ACM
  Press.

\bibitem[SS02b]{interactionRS}
K.~Swearingen and R.~Sinha.
\newblock {Interaction Design for Recommender Systems}.
\newblock In {\em Proceedings of the Conference on Designing Interactive
  Systems (DIS'02)}, London, England, June 2002. ACM Press.

\bibitem[Ste00]{decomSum}
G.~W. Stewart.
\newblock {The Decomposition Approach To Matrix Computation}.
\newblock {\em IEEE/AIP Computing in Science and Engineering}, Vol. 2(1):pages
  50--58, January--February 2000.

\bibitem[SV99]{InformationRules}
C.~Shapiro and H.~R. Varian.
\newblock {\em {Information Rules: A Strategic Guide to the Network Economy}}.
\newblock Harvard Business School Press, November 1999.

\bibitem[SW93]{socialNet}
M.~F. Schwartz and D.~C.~M. Wood.
\newblock {Discovering Shared Interests Using Graph Analysis}.
\newblock {\em Communications of the ACM}, Vol. 36(8):pages 78--89, August
  1993.

\bibitem[SWY75]{vector-space}
G.~Salton, A.~Wong, and C.~S. Yang.
\newblock {A Vector Space Model for Automatic Indexing}.
\newblock {\em Communications of the ACM}, Vol. 18(11):pages 613--620, November
  1975.

\bibitem[SYV01]{commBasedServiceLoc}
M.~P. Singh, B.~Yu, and M.~Venkatraman.
\newblock {Community-Based Service Location}.
\newblock {\em Communications of the ACM}, Vol. 44(4):pages 49--54, April 2001.

\bibitem[TH02]{HCIbookRS}
L.~Terveen and W.~Hill.
\newblock {Human-Computer Collaboration in Recommender Systems}.
\newblock In J.~M. Carroll, editor, {\em {Human-Computer Interaction in the New
  Millennium}}, {Chapter}~22. Addison-Wesley, 2002.

\bibitem[THA{\etalchar{+}}97]{PHOAKS}
L.~Terveen, W.~Hill, B.~Amento, D.~McDonald, and J.~Creter.
\newblock {PHOAKS: A System for Sharing Recommendations}.
\newblock {\em Communications of the ACM}, Vol. 40(3):pages 59--62, March 1997.

\bibitem[TOC03]{TOCHI-RS}
{\em ACM Transactions on Computer-Human Interaction}, 2003.
\newblock Special issue on recommender system interfaces: theory and practice.
  To appear.

\bibitem[Wat99]{kevinBacon}
D.~J. Watts.
\newblock {Kevin Bacon, the Small-World, and Why It All Matters}.
\newblock {\em Santa Fe Institute Bulletin}, Vol. 14(2), 1999.

\bibitem[Wel01]{compNetAsSNs}
B.~Wellman.
\newblock {Computer Networks As Social Networks}.
\newblock {\em Science}, Vol. 293:pages 2031--2034, September 2001.

\bibitem[WF94]{SNA}
S.~Wasserman and K.~Faust.
\newblock {\em {Social Network Analysis: Methods and Applications}}.
\newblock Cambridge University Press, New York, 1994.

\bibitem[WG94]{AdvancesSNA}
S.~Wasserman and J.~Galaskiewicz, editors.
\newblock {\em {Advances in Social Network Analysis: Research In The Social And
  Behavioral Sciences}}.
\newblock Sage, Thousand Oaks, CA, 1994.

\bibitem[WM99]{footprints}
A.~Wexelblat and P.~Maes.
\newblock {Footprints: History-Rich Tools for Information Foraging}.
\newblock In {\em Proceedings of the ACM Conference on Human Factors in
  Computing Systems (CHI'99)}, pages 270--277, Pittsburgh, PA, May 1999. ACM
  Press.

\bibitem[WPB01]{machineLearningForUserModeling}
G.~I. Webb, M.~J. Pazzani, and D.~Billsus.
\newblock {Machine Learning for User Modeling}.
\newblock {\em User Modeling and User-Adapted Interaction}, Vol. 11:pages
  19--29, 2001.

\bibitem[WS98]{WSsmallWorld}
D.~J. Watts and S.~Strogatz.
\newblock {Collective Dynamics of `Small-World' Networks}.
\newblock {\em Nature}, Vol. 393:pages 440--442, June 1998.

\bibitem[YGM99]{SIFT}
T.~W. Yan and H.~Garc{\'{i}}a-Molina.
\newblock {The {SIFT} Information Dissemination System}.
\newblock {\em ACM Transactions on Database Systems}, Vol. 24(4):pages
  529--565, December 1999.

\bibitem[ZK02]{trustRS}
J.~Zimmerman and K.~Kurapati.
\newblock {Exposing Profiles to Build Trust in a Recommender}.
\newblock In {\em Proceedings of the ACM Conference on Human Factors in
  Computing Systems (CHI'02)}, pages 608--609, Minneapolis, MN, April 2002. ACM
  Press.

\end{thebibliography}

\end{article}

\end{document}